\documentclass[a4paper,11pt]{article}

\pdfoutput=1

\usepackage{jheppub}
\usepackage[utf8]{inputenc}
\usepackage[T1]{fontenc}
\usepackage[capitalize]{cleveref}
\usepackage{esdiff}
\usepackage{booktabs}
\usepackage{mathtools}
\usepackage{graphicx}

\newcommand{\IE}{i.\,e.}
\newcommand{\EG}{e.\,g.}
\renewcommand{\Re}{\text{Re}}
\renewcommand{\Im}{\text{Im}}
\newcommand{\SUtwoL}{\ensuremath{{{SU(2)}_L}}}
\newcommand{\SUthreec}{\ensuremath{{{SU(3)}_c}}}
\newcommand{\sTop}{\ensuremath{\tilde{t}}}
\newcommand{\sBot}{\ensuremath{\tilde{b}}}
\newcommand{\sTau}{\ensuremath{\tilde{\tau}}}
\newcommand{\GeV}{\ensuremath{\mathrm{GeV}}}
\newcommand{\TeV}{\ensuremath{\mathrm{TeV}}}
\newcommand{\stablec}{dark green}
\newcommand{\longlc}{light green}
\newcommand{\medlc}{yellow}
\newcommand{\shortlc}{red}

\newcommand{\sTopvc}{yellow}
\newcommand{\sBotvc}{blue}
\newcommand{\sTauvc}{red}
\newcommand{\sTopsBotvc}{green}
\newcommand{\sTopsTauvc}{orange}
\newcommand{\sBotsTauvc}{purple}

\title{Impact of Vacuum Stability Constraints on the Phenomenology of
        Supersymmetric Models}

\author[a,b]{Wolfgang G.\ Hollik,}
\affiliation[a]{
Institute for Nuclear Physics (IKP), Karlsruhe Institute of Technology, \\
Hermann-von-Helmholtz-Platz 1, D-76344 Eggenstein-Leopoldshafen, Germany \\
Institute for Theoretical Particle Physics (TTP), Karlsruhe Institute of Technology, \\
Engesserstra{\ss}e 7, D-76128 Karlsruhe, Germany}
\author[b]{Georg Weiglein,}
\author[b]{Jonas Wittbrodt}
\affiliation[b]{DESY, Notkestraße 85, 22607 Hamburg, Germany}

\emailAdd{hollik@kit.edu}
\emailAdd{georg.weiglein@desy.de}
\emailAdd{jonas.wittbrodt@desy.de}

\preprint{
	\begin{minipage}[t]{0.3\textwidth}
		\begin{flushright}
			DESY 18--148 \\
			TTP 18--036
		\end{flushright}
	\end{minipage}
	}

\abstract{We present a fast and efficient method for studying vacuum stability
  constraints in multi-scalar theories beyond the Standard Model. This method is
  designed for a reliable use in large scale parameter scans. The minimization
  of the scalar potential is done with the well-known polynomial homotopy
  continuation, and the decay rate of a false vacuum in a multi-scalar theory
  is estimated by an exact solution of the bounce action in the one-field case.
  We compare to more precise calculations of the tunnelling path at the tree-
  and one-loop level and find good agreement for the resulting constraints on
  the parameter space. Numerical stability, runtime and reliability are
  significantly improved compared to approaches existing in the literature. This
  procedure is applied to several phenomenologically interesting benchmark
  scenarios defined in the Minimal Supersymmetric Standard Model.
  We utilize our efficient approach to study the impact of
  simultaneously varying multiple fields and illustrate the importance of
  correctly identifying the most dangerous minimum 
  among the minima that are deeper than the
  electroweak vacuum.
}

\begin{document}
\maketitle
\flushbottom

\section{Introduction}
\label{sec:intro}

In the Standard Model (SM) the vacuum state is characterised by a non-zero
vacuum expectation value (vev) of the Higgs field arising from the postulated
form of the Higgs potential. This vacuum breaks the electroweak (EW) symmetry
and gives masses to the particles via the Brout--Englert--Higgs mechanism. In
the standard model of big bang cosmology, the phase transition into this EW
vacuum happens within the first instants of the existence of the universe.  As
the EW vacuum is formed at that time, its stability is therefore required on
time scales of the lifetime of the universe.

In the SM the EW vacuum is stable at the EW scale by
construction. However, when extrapolating the model to high scales
this behaviour can change as a consequence of the running of the
quartic Higgs coupling $\lambda$~\cite{Sher:1988mj, Casas:1994qy,
	Espinosa1995, Isidori2001, Espinosa2008, Ellis2009, Degrassi:2012ry,
	Bezrukov:2012sa, Chetyrkin:2013wya, Buttazzo:2013uya,
	Andreassen:2014gha, Lalak2014, Zoller:2014cka, Bednyakov:2015sca,
	Branchina2015, Iacobellis2016a, Andreassen:2017rzq, Chigusa2017,
Chigusa2018}. Based on the present level of the theoretical
predictions and the experimental inputs on the top-quark and the
Higgs-boson mass, an instability occurs at scales~$\gtrsim
10^{10}\,\text{GeV}$ with a lifetime that is significantly larger than
the age of the universe. Such a configuration where the lifetime of
the vacuum state is larger than the age of the universe is called
``metastable'' and is considered to be theoretically valid.  The
determination of the lifetime of the metastable vacuum follows the
procedure developed in~\cite{Coleman1977, Callan1977}.  The presence
of additional scalar degrees of freedom, as predicted in many models
of physics beyond the SM (BSM), is in principle expected to improve
the stability of the potential at high scales through its impact on
the beta function of the quartic Higgs coupling.  On the other hand,
additional scalar degrees of freedom have the potential to destabilise
the vacuum already at the EW scale by introducing additional minima of
the scalar potential.

In BSM theories the assessment of vacuum stability at the EW scale
places important constraints on the parameter space of the model under
consideration.  A sufficient condition to ensure vacuum stability at
the EW scale is to require the EW vacuum to be the global minimum
(\IE~\emph{true vacuum}) of the scalar potential. In this case, the EW
vacuum as the state with lowest potential energy is favoured and thus
absolutely stable. In case the EW vacuum is a local minimum
(\emph{false vacuum}), the corresponding parameter region may still be
considered as allowed if it is metastable. On the contrary, any
configuration that predicts a shorter lifetime than the age of the
universe is considered to be unstable and thus excluded as it is
inconsistent with the observed lifetime of the universe.

In rather simple models, such as the Two-Higgs-Doublet Model, analytic
conditions for absolute stability have been derived~\cite{Ferreira2004,
	Barroso2006, Barroso2006a, Maniatis2006, Barroso2007, Barroso:2013awa,
Barroso2013}. In theories with more scalars analytic approaches can still be
applied to a simplifying subset of fields. However, conclusions about vacuum
stability can change severely when additional degrees of freedom are considered
(see e.g.~\cite{Muhlleitner:2016mzt}). Thus, numerical approaches that can
account for a variety of fields simultaneously are of interest. Supersymmetry
(SUSY) requires an extended Higgs sector as compared to the SM case and adds a
scalar degree of freedom for every fermionic one in the SM\@. Therefore even the
Minimal Supersymmetric Model (MSSM) corresponds to a multi-scalar theory with a
much richer scalar sector than the SM\@. Many different approaches have been
employed in order to obtain constraints from vacuum stability for the
MSSM~\cite{Frere:1983ag, Claudson:1983et, Drees:1985ie, Gunion:1987qv,
	Komatsu:1988mt, Langacker:1994bc, Casas:1995pd, Strumia:1996pr, Kusenko:1996jn,
	Casas:1996de, LeMouel:1997tk, Abel:1998wr, LeMouel:2001sf, Ferreira:2001tk,
	Ferreira:2000hg, Ferreira:2004yg, Park:2010wf, Altmannshofer:2012ks, Carena:2012mw,
	Camargo-Molina:2013sta, Chattopadhyay:2014gfa, Camargo-Molina2014,
	Chowdhury:2013dka, Blinov:2013fta, Bobrowski:2014dla, Altmannshofer:2014qha,
Hollik:2015pra, Hollik:2016dcm}.

Public tools that allow one to efficiently obtain constraints from vacuum
stability in general BSM models would clearly be very useful in this context. To
our knowledge \texttt{Vevacious}~\cite{Camargo-Molina2013, Camargo-Molina2014},
which is designed to check the stability of the EW vacuum including
\mbox{one-loop} and finite temperature effects, is the only dedicated public
tool that is applicable to a variety of BSM models.\footnote{The code
	\texttt{BSMPT}~\cite{Basler:2018cwe}---while not designed for vacuum stability
	studies---can also check for absolute stability in several BSM models including
one-loop and finite temperature effects.} In this paper we present an approach
that provides a highly efficient and reliable evaluation of the constraints from
vacuum stability such that they can be incorporated into BSM parameter scans,
which typically run over a large number of points in a multi-dimensional
parameter space.  Our approach is applicable to any model with a renormalisable
scalar potential and has been validated on SUSY and non-SUSY models.  In this
work we outline our method and, as an example, apply it to the MSSM benchmark
scenarios for Higgs searches at the LHC that were recently defined
in~\cite{Bahl2018a}.  These benchmark scenarios were designed to provide
interesting Higgs phenomenology at the LHC\@. Limits from searches for SUSY
particles as well as other constraints affecting the MSSM Higgs sector were
taken into account in the definitions of the benchmark scenarios, while no
detailed investigation of the parameter planes of the scenarios with respect to
constraints from vacuum stability has been carried out.  We perform such an
analysis of the various benchmark planes and indicate the parameter regions that
are incompatible with the requirement of a sufficiently stable EW vacuum. A
public tool based on the implementation of our method with which the numerical
results in this paper have been obtained is in preparation.

This paper is organised as follows: we begin with the description of our method
in \cref{sec:deepdir} by discussing the most general renormalisable scalar
potential in a form that is suitable for vacuum stability calculations and
compare different methods for estimating the vacuum lifetime. In
\cref{sec:susy}, we fix our notation for the MSSM and the relevant field space.
Finally, we apply our method to the MSSM in \cref{sec:bench} and discuss in
detail the results for three of the benchmark scenarios defined
in~\cite{Bahl2018a}. We conclude in \cref{sec:conclusion}. Furthermore,
\cref{app:pot} contains the full MSSM scalar potential and \cref{app:rest}
illustrates our results for the remaining CP-conserving benchmark scenarios
proposed in~\cite{Bahl2018a}.

\section{Vacuum Stability in Multidimensional Field Spaces}
\label{sec:deepdir}
The vacuum state of a (quantum) field theory is determined by the
state of lowest potential energy. In field theory, this state is the
minimum of the (effective) potential \(V(\phi)\), which describes the
potential energy density of a field \(\phi\). In general, the
Lagrangian for such a real scalar field \(\phi\) is given by
\begin{equation}\label{eq:scalLag}
	\mathcal{L} = \frac{1}{2} {\left(\partial \phi \right)}^2 - V(\phi),
\end{equation}
where \(V\) can be an arbitrary function of the field \(\phi\) that is
bounded from below. In a renormalisable quantum field theory at
tree-level, it may contain all interactions up to quartic
terms. Formally, the effective potential is defined for
\emph{classical} field values \(\phi_\text{cl}\) that minimise the
effective action. For our purpose, the field theoretical potential and
the effective potential are the same when replacing field operators
\(\phi\) by classical commuting field values \(\phi_\text{cl}\) and
defining the effective potential \(V(\phi = \phi_\text{cl})\) as
function of \(\phi \equiv
\phi_\text{cl}\)~\cite{Coleman:1973jx}. Thus, we treat all scalar
fields as commuting variables.

We consider now the general case of \(n\) real scalar fields $\phi_a$
with $a\in\lbrace 1,\ldots,n\rbrace$ in a renormalisable quantum field
theory at tree-level
\begin{equation}
	V(\vec\phi) = \lambda_{abcd}\phi_a\phi_b\phi_c\phi_d
	+ A_{abc}\phi_a\phi_b\phi_c
	+ m^2_{ab}\phi_a\phi_b + t_{a}\phi_a + c\,,\label{eq:genpot}
\end{equation}
where the sum over repeated indices is implied. The totally symmetric
coefficient tensors $\lambda_{abcd}$, $A_{abc}$, $m^2_{ab}$ and $t_a$
as well as the constant $c$ contain all possible real coefficients
with non-negative mass dimension. This potential includes in general
up to \(3^n\) stationary points\footnote{Note that we discard complex
solutions here as we consider real fields.} out of which an initial
vacuum at $\vec\phi=\vec\phi_v$ is selected as minimum with
\begin{equation}
	\left.\diffp{ V}{ {{\phi_a}}}\right|_{\vec\phi=\vec\phi_v}=0\,,\label{eq:firstderivzero}
\end{equation}
and the mass matrix
\begin{equation}
	M_{ab}=\left.\diffp{ V}{{{ \phi_a}} {{\phi_b}}}\right|_{\vec\phi=\vec\phi_v}\label{eq:posmassm}
\end{equation}
is positive definite. After expanding \cref{eq:genpot} around the vacuum
as $\vec\phi=\vec\phi_v+\vec\varphi$, with \(\vec\varphi = {(
	\varphi_1, \ldots, \varphi_n )}^\mathrm{T}\), we obtain
\begin{equation}
	V(\vec\varphi)=\lambda'_{abcd}\varphi_a\varphi_b\varphi_c\varphi_d
	+ A'_{abc}\varphi_a\varphi_b\varphi_c
	+ m'^2_{ab}\varphi_a\varphi_b\,,\label{eq:expandingpot}
\end{equation}
where $t'_a$ vanishes due to \cref{eq:firstderivzero}, and we have
normalised the potential energy at $\vec\varphi=0$ to zero. For
particle physics applications, this normalisation plays no role. Note,
however, that a constant term yields a non-vanishing cosmological
constant~\cite{Weinberg:1988cp}. We rewrite the field-space vector as
$\vec\varphi\rightarrow \varphi \hat\varphi$ with a unit vector
\(\hat\varphi\) and its absolute value \(\varphi = \sqrt{\varphi_1^2 +
	\ldots + \varphi_n^2}\) and obtain
\begin{equation}
	V(\varphi,\hat\varphi)=\lambda(\hat\varphi)\varphi^4
	- A(\hat\varphi)\varphi^3 + m^2(\hat\varphi)\varphi^2\,, \label{eq:Vdir}
\end{equation}
where all the dependence on the normalised direction in field space
$\hat\varphi$ has been absorbed into the coefficients $\lambda$, $A$ and $m^2$.
The potential has to be bounded from below, so $\lambda>0$ for all directions
$\hat\varphi$. Furthermore, in order to have a minimum at \(\varphi = 0\), the
condition $m^2>0$ has to be satisfied. There is a freedom of sign choice in
either \(\varphi\) or \(A\). We always choose \(A > 0\) without loss of
generality and have defined the minus sign in \cref{eq:Vdir} such that a
possible minimum will be located in $\varphi>0$.

\begin{figure}
	\centering
	\includegraphics[width=0.8\textwidth]{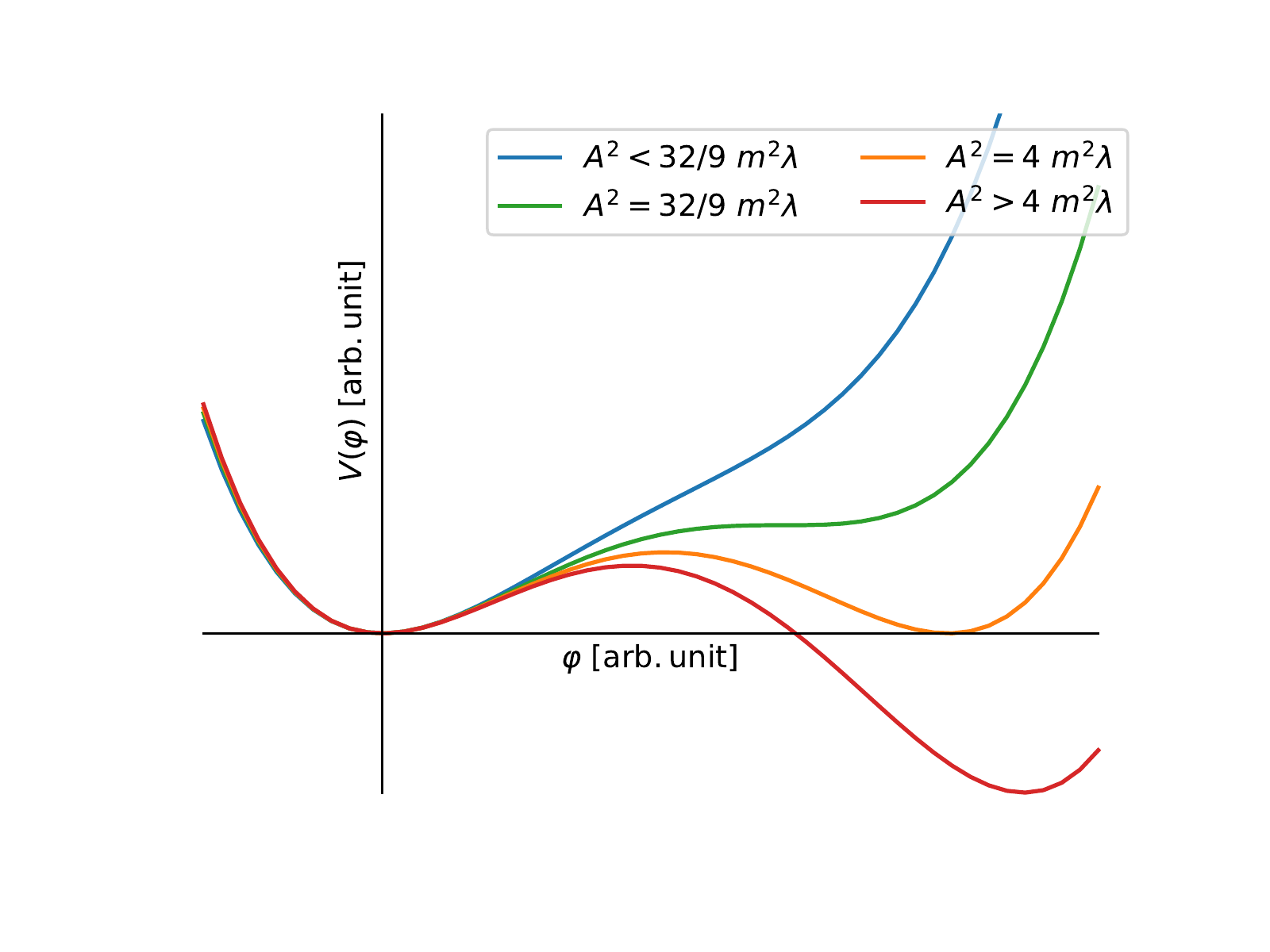}
	\caption{Behaviour of the generic quartic potential as given in
		\cref{eq:Vdir} for different relations between the coefficients
		\(A\), \(m^2\) and \(\lambda\) as indicated in the
		legend for arbitrary units on both axes.}\label{fig:possibilities}
\end{figure}

\Cref{fig:possibilities} shows the resulting possible shapes of the
potential in \cref{eq:Vdir}. Since it is a quartic polynomial in one
variable it can have at most two minima, one of which we have chosen to
lie at the origin. The second minimum exists as soon as
\begin{equation}
	{\left(A(\hat\varphi)\right)}^2>\frac{32}{9}m^2(\hat\varphi)\lambda(\hat\varphi)
\end{equation}
and is deeper than the minimum at the origin if
\begin{equation}
	{\left(A(\hat\varphi)\right)}^2 > 4 m^2(\hat\varphi) \lambda(\hat\varphi)\,. \label{eq:deepdir}
\end{equation}
This discussion implies that large cubic terms $A$ compared to the mass
parameters and self-couplings are potentially dangerous for the stability of the
initial vacuum at the origin. We call the directions $\hat\varphi$
fulfilling \cref{eq:deepdir} \emph{deep directions}.

This simple form is very useful for the calculation of vacuum decay in
\cref{sec:bounce}. However, many disjoint regions of deep directions may exist
which makes the numerical search for such directions on the unit $(n-1)$-sphere
of directions $\hat\varphi$ infeasible, see \EG{} Ref.~\cite{Hollik:2016dcm}. We
instead use the numerical method of polynomial homotopy continuation (PHC) (see
\EG{}~\cite{morgan2009solving} or~\cite{Maniatis:2012ex}) to find all stationary
points of \cref{eq:genpot}. From these stationary points we select the deep
directions by comparing their depth to the initial vacuum.

PHC efficiently finds all solutions of systems of polynomial equations. We use
it to solve
\begin{equation}
	\vec\nabla_\phi V = 0
\end{equation}
and find all real solutions, \IE{} the stationary points of the scalar
potential. While PHC in theory never fails to find all solutions of the system,
solutions may be missed due to numerical uncertainties in judging whether a
solution is real or complex. This can be avoided by a careful preconditioning of
the system of equations~\cite{morgan2009solving}. Another subtlety is that PHC
only finds point-like, isolated solutions. This is especially important in the
physically interesting cases of gauge theories where any vacuum is only unique
up to gauge transformations. If any gauge freedom is left in the model, this
turns all isolated solutions into continuous curves which cannot be found by the
algorithm. For this reason it is essential to implement models with all gauge
redundancies removed. For the case of at least one Higgs doublet this can be
achieved by setting the charged and imaginary components of one Higgs doublet to
zero without loss of generality.

\subsection{Calculation of the Bounce Action}\label{sec:bounce}
We briefly review the definition of the so-called bounce action, which describes
the decay of a false vacuum. Consider a single real field Lagrangian as in
\cref{eq:scalLag}. The semi-classical tunnelling and first quantum corrections
were calculated in~\cite{Coleman1977,Callan1977}. It was found that the decay
rate $\Gamma$ of a metastable vacuum state per (spatial)
volume $V_{S}$ is given by the
exponential decay law
\begin{equation}
	\frac{\Gamma}{V_{S}}=K e^{-B}\,,\label{eq:decaydef}
\end{equation}
where \(K\) is a dimensionful parameter that will be specified below, and \(B\)
denotes the bounce action which gives the dominant contribution to \(\Gamma\).
The bounce $\phi_B(\rho)$ is the solution of the euclidean equation of motion
\begin{equation}
	\diff[2]{\phi}{\rho}+\frac{3}{\phi}\diff{\phi}{\rho}=\diffp{U}{\phi}\label{eq:bouncediff}
\end{equation}
with the boundary conditions
\begin{equation}
	\phi(\infty)=\phi_v\,, \quad \left.\diff{\phi}{\rho}\right|_{\rho=0}=0\,.
\end{equation}
$U$ is the euclidean scalar potential, $\rho$ is a spacetime variable and
$\phi_v$ is the location of the metastable minimum. The bounce action $B$ is the
stationary point of the euclidean action given by the integral
\begin{equation}
	B = 2\pi^2\int^\infty_0 \rho^3\text{d}\rho \left[\frac{1}{2}{\left(\diff{}{\rho}\phi_B(\rho)\right)}^2 + U(\phi_B(\rho))\right]\,.\label{eq:bounceint}
\end{equation}

In the one field case, \cref{eq:bounceint,eq:bouncediff} can be solved
numerically by the undershoot/overshoot method (see e.g.~\cite{Apreda:2001us}).
While all of the above equations generalise trivially to the multi-field case
$\phi\rightarrow\vec\phi$, the strategy for obtaining the decay rate becomes
considerably more involved. In order to judge the stability of the EW vacuum we
need to obtain the minimal bounce action for tunnelling into a deeper point in
the scalar potential. There exist methods for solving \cref{eq:bouncediff}
numerically in multiple field dimensions~\cite{Kusenko:1996jn, Dasgupta:1996qu,
Moreno:1998bq, Cline:1999wi, John:1998ip, Cline:1998rc, Konstandin:2006nd,
Park:2010rh, Wainwright:2011kj, Masoumi:2016wot,Athron:2019nbd} using
optimization, discretisation, path-deformation or multiple shooting. As we will
show in the next section a fast evaluation of the bounce action is more
important for our purposes than an extremely precise result. For this reason, we
approximate the path of the bounce by the straight line in a given deep
direction. The potential along this straight path is a simple quartic polynomial
as given by \cref{eq:Vdir}. For this form of the potential there exists a
semi-analytic result for the bounce action~\cite{Adams:1993zs}
\begin{equation}
	B = \frac{\pi^2}{3\lambda}{(2-\delta)}^{-3}\left(13.832\,\delta-10.819\,\delta^2+2.0765\,\delta^3\right)\label{eq:bounceapp}
\end{equation}
with
\begin{equation}
	\delta=\frac{8\lambda m^2}{A^2}\,.
\end{equation}
The expression in brackets was obtained in~\cite{Adams:1993zs} by fitting a
cubic polynomial in $\delta$ to the numerical result. The coefficients do not
depend on any model parameters, and the polynomial approximation agrees with the
numerical result within a $0.004$ absolute tolerance for all values of $\delta$.
We use this formula to calculate $B$ for all deep directions from the initial
vacuum. The deep direction with the smallest bounce action is the dominant decay
path.

The value of $B$ is not invariant under rescaling of the field $\varphi$
of \cref{eq:Vdir}
\begin{align}
	\varphi\to n \varphi & \Rightarrow \lambda\to n^4\lambda\,,\ A\to n^3 A\,,\ m^2\to n^2 m^2\,,\ \delta\to\delta \\
	                     & \Rightarrow B\to n^{-4}B\,.                                                             
\end{align}
This dependence on the field normalisation arises from the equation of
motion, where \cref{eq:bouncediff} only applies to fields with canonically
normalised kinetic terms. A consistent expansion of the form of \cref{eq:Vdir}
therefore requires all real field components to have canonically normalised
kinetic terms. It is crucial to ensure that the implementation of the scalar
potential fulfils this requirement.

\subsection{Lifetime of the Metastable Vacuum}\label{sec:lifetime}
The vacuum lifetime in \cref{eq:decaydef} also depends on the quantity $K$. The
value of $K$ is both challenging to calculate and a subdominant effect towards
the tunnelling rate as it does not enter in the exponent. Since it is a
dimensionful parameter, $[K]=\GeV^4$, it can be estimated from a typical scale
$\mathcal{M}$ of the theory as
\begin{equation}
	K = \mathcal{M}^4\,.
\end{equation}
Comparing the vacuum decay time \(\tau_\text{decay}\) with the age of
the universe \(t_\text{uni}\)~\cite{Aghanim:2018eyx} yields~\cite{Andreassen:2017rzq}
\begin{align}
	\frac{\tau_\text{decay}}{t_\text{uni}}                                    
	= {\left(\frac{\Gamma}{V_S}\right)}^{-\frac{1}{4}} \frac{1}{t_\text{uni}} 
	= \frac{1}{t_\text{uni}\mathcal{M}}e^{B/4}\,. \label{eq:relStab}          
\end{align}

\begin{figure}
	\centering
	\includegraphics[width=0.8\textwidth]{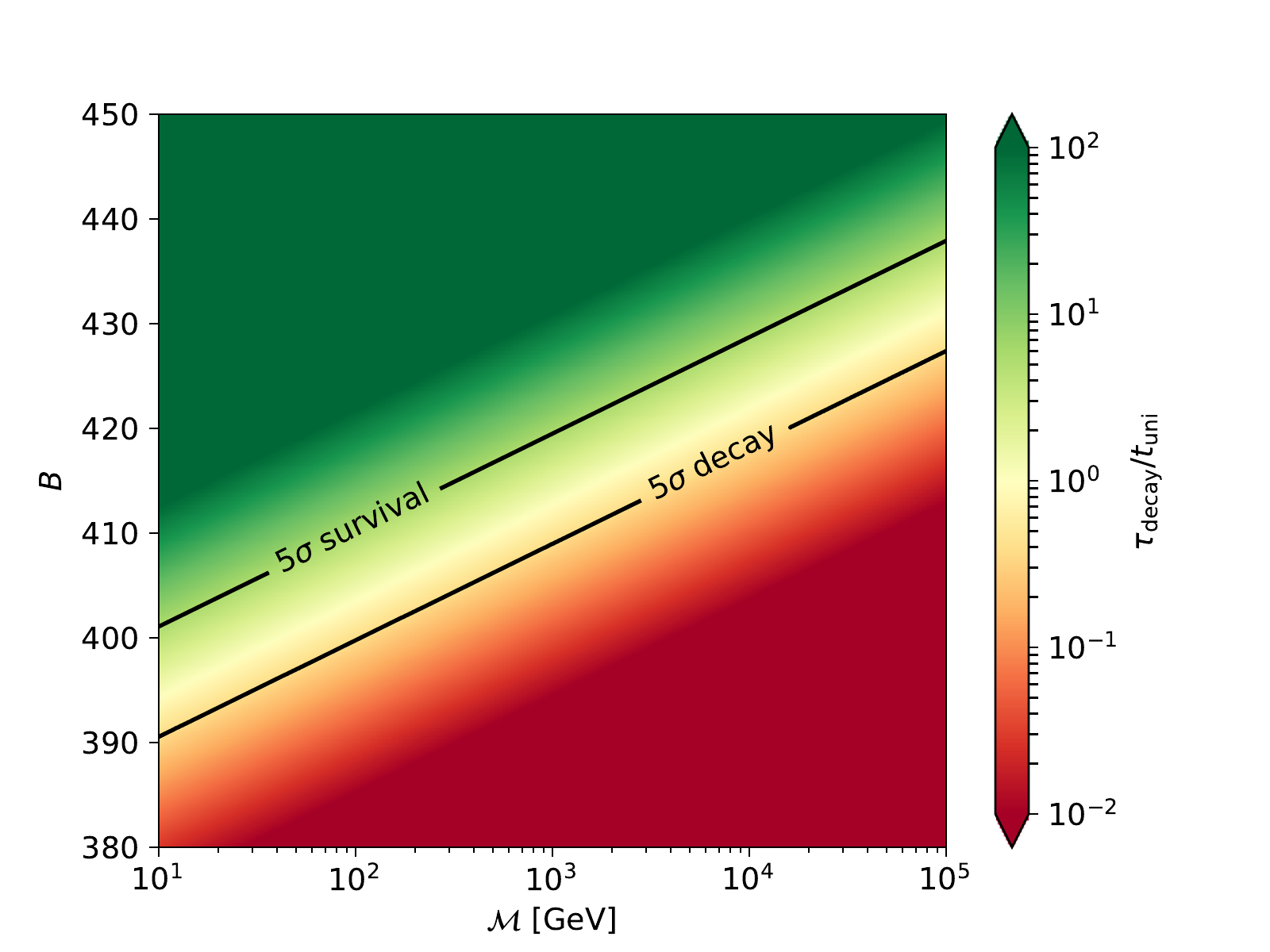}
	\caption{The lifetime of the metastable vacuum $\tau_\text{decay}$ relative to
		the age of the universe $t_\text{uni}$ is given in the plane of the scale $\mathcal{M}$ and
		the bounce action $B$. The contour lines denote a $5\sigma$ probability for
		decay and survival, respectively.}\label{fig:lifetime}
\end{figure}

\Cref{fig:lifetime} shows the relative lifetime $\tau_\text{decay}/t_\text{uni}$
as a function of $B$ and $\mathcal{M}$. As expected, the threshold of
instability where $\tau_\text{decay}\sim t_\text{uni}$ is highly sensitive to
$B$ and only mildly sensitive to $\mathcal{M}$. In \cref{fig:lifetime} we also
show the contours corresponding to a $5\,\sigma$ expected decay or a $5\,\sigma$
expected survival of the vacuum
during the evolution of the universe. The survival probability
is given by
\begin{equation}
	P = \exp\left(-\frac{\Gamma}{V_S}\tilde{V}_\text{light-cone}\right) = \exp\left(- \mathcal{M}^4
	\tilde{V}_\text{light-cone}e^{-B}\right)\,,
\end{equation}
where the (spacetime) volume of the past light-cone is
$\tilde V_\text{light-cone}\sim 0.15/H_0^4$~\cite{Buttazzo:2013uya}, and
$H_0$ is the current value of the Hubble parameter~\cite{Aghanim:2018eyx}.
The points in the
green region of \cref{fig:lifetime} are definitely long-lived with respect to
the age of the universe, while the red points are definitely short-lived. We see
that varying the scale $\mathcal{M}$ over a generous range from $10\,\GeV$ to
$100\,\TeV$ shifts the border between metastability and instability by less than
$10\%$ in $B$. We therefore consider any point where $B>440$ as long-lived and
any point where $B<390$ as short-lived. We treat the
intermediate range $390<B<440$ as an uncertainty on the stability threshold from
the unknown $\mathcal{M}$.

\subsection{Parameter Scans and the Reliability of Vacuum Stability
	Calculations}\label{sec:precision}

Our main objective is to enable the use of vacuum stability constraints
in fits and parameter scans of BSM theories. Typical BSM parameter scans
consider millions of different parameter points which places strong
requirements on evaluation time and reliability.

The first trade-off between speed and precision is in the calculation
of the bounce action described in \cref{sec:bounce}. Instead of using
one of the available sophisticated solvers~\cite{Wainwright:2011kj,
Masoumi:2016wot}---which may take a lot of runtime and could
encounter numerical problems---we approximate the tunnelling path with
a straight line in field space and use the semi-analytic solution
\cref{eq:bounceapp}~\cite{Adams:1993zs}. A comparison between such
simple approximations and the multiple-shooting method
of~\cite{Masoumi:2016wot} has been performed in~\cite{Masoumi:2017trx}
where agreement within $\mathcal{O}(10\%)$ for polynomial potentials
has been found. The approximation according to \cref{eq:bounceapp} is
evaluated instantaneously while the available solvers typically take
between a few seconds and several minutes per tunnelling calculation.

This approximation, as well as the discussion of \cref{sec:deepdir},
relies on the potential being a quartic polynomial which is only true at
tree-level. In contrast, vacuum stability in the SM has been studied up
to full NNLO~\cite{Sher:1988mj, Casas:1994qy, Degrassi:2012ry,
	Buttazzo:2013uya, Andreassen:2014gha, Bednyakov:2015sca,
Andreassen:2017rzq} precision involving the two-loop effective Higgs
potential and NNLO running and threshold effects.  Even for BSM models
the public tool \texttt{Vevacious}~\cite{Camargo-Molina2013,
Camargo-Molina2014} can perform vacuum stability calculations using
the one-loop Coleman--Weinberg potential.  However, it has recently been
shown~\cite{Andreassen:2016cff, Andreassen:2016cvx} that the use of the
loop-corrected effective potential for stability calculations is not in
general a consistent perturbative expansion. This happens because the
effective action is a perturbative expansion in both the usual powers of
$\hbar$ and the momentum transfer, where the effective potential
corresponds to the zeroth order term of the momentum
expansion. Truncating this second expansion does not in general provide
a good approximation in calculations of the bounce action. Therefore,
higher momentum terms of the full effective action can give
contributions to the bounce action comparable to the contributions from
the effective potential. Since it seems unfeasible to calculate even the
leading higher momentum terms of the effective action in general BSM
models it appears questionable whether using the one-loop effective
potential for stability calculations leads to more precise results.  For
this reason we stick to the tree-level potential which allows us to
apply the very concise formulation of \cref{sec:deepdir,sec:bounce} and
considerably increases the speed and numerical stability of our
calculation.

The tunnelling time with respect to the age of the universe as given
in \cref{eq:relStab} depends exponentially on the value of the bounce
action $B$.  For any given parameter point, small uncertainties on $B$
are therefore amplified to large uncertainties on the tunnelling
time. While this makes precise predictions for the lifetime of
individual parameter points very challenging, it is less problematic
for constraining the parameter space of BSM models since the bounce
action $B$ is also very sensitive to the values of the model
parameters. Therefore, a small shift in parameter space typically
leads to a change in the bounce action substantially larger than the
uncertainties described above. For this reason the resulting
constraints on the model parameter space depend only mildly on the
precise way $B$ is calculated.  This dependence can be estimated from
the width of the uncertainty band $390<B<440$ (see
\cref{sec:lifetime}) that we will show in all of our results (see
e.g.\ \cref{fig:Mh125tri,fig:Mh125alitri} below).

\section{Application to the Minimal Supersymmetric Standard Model}
\label{sec:susy}

Constraints from vacuum stability can have an important impact on
supersymmetric models. The main reason is their abundance of scalar
fields with at least two Higgs doublets as well as two scalar
superpartners for each SM fermion. The MSSM is the simplest and best
studied supersymmetric model where only a second Higgs doublet is
added to the Higgs sector~\cite{Nilles:1983ge, Haber:1984rc,
Gunion:1984yn}. The most important vacuum stability constraints in
the MSSM concern instabilities in directions with sfermion vevs. These
are commonly referred to as charge or colour breaking (CCB)
vacua~\cite{Frere:1983ag, Claudson:1983et, Drees:1985ie,
	Gunion:1987qv, Komatsu:1988mt, Langacker:1994bc, Casas:1995pd,
	Strumia:1996pr, Kusenko:1996jn, Casas:1996de, LeMouel:1997tk,
	Abel:1998wr, LeMouel:2001sf, Ferreira:2001tk, Ferreira:2000hg,
	Ferreira:2004yg, Park:2010wf, Altmannshofer:2012ks, Carena:2012mw,
	Camargo-Molina:2013sta, Chattopadhyay:2014gfa, Camargo-Molina2014,
	Chowdhury:2013dka, Blinov:2013fta, Bobrowski:2014dla,
Altmannshofer:2014qha, Hollik:2015pra, Hollik:2016dcm}. The existing
results include both analytic and semi-analytic studies of specific
directions in field space as well as fully numeric approaches. After
specifying our notation for the scalar potential of the MSSM we are
going to illustrate our treatment of vacuum stability constraints for
an example application to MSSM benchmark scenarios for Higgs searches.

\subsection{Treatment of the MSSM Scalar Potential}\label{sec:MSSMpot}
For our discussion, we focus on the third generation of SM fermions and their
corresponding superpartners as those have the largest couplings to the Higgs
sector. We comment on the impact of the sfermions of the first and second
generations below. The superpotential of the MSSM including only third
generation fermion superfields is given by
\begin{equation}
	W = \mu  H_u \cdot H_d
	+ y_t Q_L \cdot H_u \bar t_R
	+ y_b  H_d \cdot Q_L \bar b_R
	+ y_\tau H_d \cdot L_L \bar\tau_R\,,\label{eq:MSSMspot}
\end{equation}
with the chiral Higgs superfield \SUtwoL{} doublets $H_u=(H^0_d, H_d^-)$ and
$H_d = (H_u^+,H_u^0)$, the left-chiral superfields containing the SM quark
and lepton doublets $Q_L = (t_L, b_L)$ and $ L_L=(\nu_L,\tau_L)$,
respectively, as well as the superfields containing the \SUtwoL{} singlets
$\bar t_R$, $\bar b_R$ and $\bar \tau_R$. The Yukawa couplings are denoted
by $y_{t,b,\tau}$ and the dot product is the \SUtwoL{} invariant
multiplication $\Phi_i\cdot\Phi_j=\epsilon_{ab}\Phi_i^a\Phi_j^b$ with the
totally antisymmetric tensor \(\epsilon_{ab}\), where \(\epsilon_{12} =
-1\). The superpotential gives rise to the $F$-terms
\begin{equation}
	F = \sum_{\phi}\left|\partial_x W\right|^2\,, \quad \phi \in \{h_u^0,
	h_u^+,h_d^0,h_d^-,\sTop_L,\sBot_L,\sTau_L,\tilde{\nu}_L,\sTop_R^*,\sBot_R^*,\sTau_R^*\}
	\label{eq:MSSMFtems}
\end{equation}
contributing to the scalar potential, where the sum runs over all
scalar components of the superfields in \cref{eq:MSSMspot}. Note that
the $F$-terms contain quadratic, cubic and quartic terms.

Additional supersymmetric contributions to the scalar potential come
from the gauge structure of the model. These $D$-terms are given by
\begin{subequations}
	\begin{align}
		D             & = D_{{U(1)}_Y} + D_\SUtwoL{} + D_\SUthreec{}\,,\label{eq:MSSMdterms}                                                                                             \\
		D_{{U(1)}_Y}  & =\frac{g_1^2}{8}{\left(\sum_{\phi}Y_\phi|\phi|^2\right)}^2\,,\label{eq:MSSMd1terms}                                                                              \\
		D_\SUtwoL{}   & =\frac{g_2^2}{8}\sum_{\Phi_i}\sum_{\Phi_j}2(\Phi_i^\dagger\Phi_j)(\Phi_j^\dagger\Phi_i) - (\Phi_i^\dagger\Phi_i)( \Phi_j^\dagger\Phi_j)\,,\label{eq:MSSMd2terms} \\
		D_\SUthreec{} & =\frac{g_3^2}{6}{\left(|\sTop_L|^2-|\sTop_R|^2+|\sBot_L|^2-| \sBot_R|^2\right)}^2\label{eq:MSSMd3terms}\,,                                                       
	\end{align}
\end{subequations}
where the sum over $\phi$ runs over all scalar components as in
\cref{eq:MSSMFtems}, and $\Phi_i, \Phi_j \in\{h_u,h_d,\tilde{Q}_L,
\tilde{L}_L\}$ run over the scalar \SUtwoL{} doublets. The different
prefactor for the \SUthreec{} $D$-term arises from the sum over the
\SUthreec{} generators. The $D$-terms only contain quartic terms.

Finally, another contribution to the scalar potential of the MSSM are
the soft SUSY breaking terms
\begin{equation}
	\begin{aligned}
		V_\text{soft} & =  m_{H_u}^2 h_u^\dagger h_u + m_{H_d}^2 h_d^\dagger h_d +\left(B_\mu h_u\cdot h_d+ \text{h.c.}\right)                                                                \\
		              & \quad + m_{Q_3}^2 \tilde{Q}_L^\dagger \tilde{Q}_L + m_{L_3}^2 \tilde{L}_L^\dagger \tilde{L}_L + m_{U_3}^2 |\sTop_R|^2 + m_{D_3}^2 |\sBot_R|^2 + m_{E_3}^2 |\sTau_R|^2 \\
		              & \quad + \left(y_t A_t \sTop_R^* \tilde{Q}_L\cdot h_u  + y_b A_b \sBot_R^* h_d\cdot\tilde{Q}_L + y_\tau A_\tau \sTau_R^* h_d\cdot\tilde{L}_L + \text{h.c.}\right)\,,   
	\end{aligned}
\end{equation}
where $y_{t,b,\tau}$ are the Yukawa couplings of \cref{eq:MSSMspot}. We shall
express the soft breaking parameter $B_\mu$ via the mass \(m_A\) of the CP-odd
Higgs boson,
\begin{equation}
	B_\mu=m_{A}^2 \sin\beta\cos\beta\,,
\end{equation}
using the ratio of the vevs of the two Higgs doublets at the EW vacuum
\begin{equation}
	\tan\beta=v_u/v_d\,.
\end{equation}
The full scalar potential of the MSSM including all Higgs and third
generation sfermion fields is thus given by
\begin{equation}
	V = F + D + V_\text{soft}\,,\label{eq:MSSMpot}
\end{equation}
see Appendix~\ref{app:pot} for the explicit expression.

We have so far written $V$ in terms of complex fields, while the
method outlined in \cref{sec:bounce} relies on a reduction of the
field space to a direction parametrised by a single real scalar
field. For \cref{eq:bounceapp} to be applicable, this field also needs
to have a canonically normalised kinetic term. We ensure both
requirements by expressing $V$ exclusively through real scalar fields
with canonically normalised kinetic terms and by expanding all complex
scalar fields as
\begin{equation}
	\phi \rightarrow \frac{1}{\sqrt{2}}\Re(\phi) + \frac{i}{\sqrt{2}}\Im(\phi)\,.
\end{equation}
This ensures that $\varphi$ is canonically normalised after expanding
$\vec\varphi= \varphi \hat\varphi$ to obtain \cref{eq:Vdir} as
long as $|\hat\varphi|=1$. In this notation the EW vacuum is given by
\begin{equation}
	\Re(h_u^{0}) = v \sin\beta \,,\quad \Re(h_d^{0}) = v \cos\beta\,,
\end{equation}
where $v=\sqrt{v_u^2+v_d^2}\approx246\,\GeV$ is the SM Higgs vev.

It would be unfeasible to vary all real scalar degrees of freedom simultaneously
since the runtime of the minimisation procedure scales exponentially with the
number of fields considered.\footnote{For example, considering the set of fields
	in \cref{eq:fieldstb} yields $\sim10$ times longer runtimes than varying the two
	sets of fields
	$\lbrace\Re(h_u^{0}),\,\Re(h_d^{0}),\,\Re(\sTop_L),\,\Re(\sTop_R)\rbrace$ and
	$\lbrace\Re(h_u^{0}),\,\Re(h_d^{0}),\,\Re(\sBot_L),\,\Re(\sBot_R)\rbrace$
separately.} For our studies of the MSSM we combine all stationary points found
by varying the three sets of fields
\begin{subequations}
	\begin{gather}
		\left\lbrace\Re(h_u^{0}),\,\Re(h_d^{0}),\,\Re(\sTop_L),\,\Re(\sTop_R),\,\Re(\sBot_L),\,\Re(\sBot_R)\right\rbrace\,,\label{eq:fieldstb} \\
		\left\lbrace\vphantom{\sBot}\Re(h_u^{0}),\,\Re(h_d^{0}),\,\Re(\sTop_L),\,\Re(\sTop_R),\,\Re(\sTau_L),\,\Re(\sTau_R)\right\rbrace\,,\label{eq:fieldstl} \\
		\left\lbrace\Re(h_u^{0}),\,\Re(h_d^{0}),\,\Re(\sBot_L),\,\Re(\sBot_R),\,\Re(\sTau_L),\,\Re(\sTau_R)\right\rbrace\,.\label{eq:fieldsbl}
	\end{gather}
\end{subequations}
All sets contain the real parts of the neutral Higgs fields that participate in
EW symmetry breaking. The first set additionally contains the real \sTop{} and
\sBot{} fields, the second set the \sTop{} and \sTau{} fields and the third set
the \sBot{} and \sTau{} fields. This method will not be able to find stationary
points for which \sTop{}, \sBot{} and \sTau{} vevs are simultaneously non-zero.
The distance in field space between the EW vacuum and another minimum is
expected to increase as more fields take non-zero values at this second minimum.
We therefore neglect these configurations since the tunnelling time increases
with the field-space distance. Moreover, we found $\tilde\nu$ vevs and vevs of
the first and second sfermion generations to have no impact on the observed
constraints. Therefore we are not going to show and discuss them in detail in
the following, but we will comment on their impact below. We also do not take
charged or CP-odd Higgs fields and the imaginary parts of the sfermion fields
into account. Ignoring the CP-odd and charged Higgs directions is motivated by
the absence of any spontaneous CP or charge breaking in the Higgs sector of the
2HDM~\cite{Ferreira2004,Barroso2006} (and thus the MSSM). While non-zero charged
and CP-odd Higgs vevs can in principle develop in the presence of sfermion vevs
we found no region of parameter space where these are relevant. We neglect the
imaginary parts of the sfermions as they are not expected to add new features in
the absence of CP-violation.\footnote{Apart from the CP-violation in the CKM
matrix which does not enter our study.} Note that this discussion is specific to
the MSSM\@. In the NMSSM for example the different kinds of Higgs vevs are
expected to be more relevant~\cite{Krauss:2017nlh}.

\subsection{The Impact of Yukawa Couplings}
\label{sec:deltab}
The third generation Yukawa couplings are the largest Yukawa couplings
and sensitively depend on \(\tan\beta\) already at the tree level. Their
value is determined via the quark masses\footnote{We treat the quark
	masses as running masses at the SUSY scale. To this end we use
	\texttt{RunDec}~\cite{Chetyrkin:2000yt, Schmidt2012, Herren2018} to
	run the $\overline{\text{MS}}$ quark masses to the SUSY scale assuming
SM running.} and the vev of the Higgs doublet coupling to them at the
tree level:
\begin{equation}
	y_t^\text{tree} = \frac{\sqrt{2}m_t}{v_u} = \frac{\sqrt{2}m_t}{v \sin\beta}
	\qquad \text{and} \qquad
	y_b^\text{tree} = \frac{\sqrt{2}m_b}{v_d} = \frac{\sqrt{2}m_b}{v \cos\beta}\,.\label{eq:yukcoup}
\end{equation}
For small \(\tan\beta\), the value of the bottom Yukawa coupling is
suppressed with respect to the top Yukawa coupling, while for large
\(\tan\beta\) they become comparable.

For large \(\tan\beta\), the bottom Yukawa coupling is very sensitive to
SUSY loop corrections that are enhanced by \(\tan\beta\). The
leading corrections can be resummed~\cite{Hall:1993gn, Carena:1994bv,
Pierce:1996zz, Carena:1999py}, and it is advantageous to include them
despite the fact that the scalar potential is evaluated at the tree
level, since they effectively change the value of the bottom Yukawa
coupling. The impact of the resummed corrections on the Yukawa coupling
can be included by replacing $y_b^\text{tree}$ in \cref{eq:yukcoup} by
\begin{equation}\label{eq:botres}
	y_b^\text{res} = \frac{\sqrt{2}m_b}{v_d (1 + \Delta_b)} \,,
\end{equation}
where \(\Delta_b\) contains SUSY loop corrections. The dominant
contributions arise from the gluino-sbottom and higgsino-stop loop, which enter
in the sum \(\Delta_b = \Delta_b^\text{gluino} + \Delta_b^\text{higgsino}\):
\begin{subequations}
	\begin{align}
		\Delta_b^\text{gluino}   & = \frac{2 \alpha_s}{3\pi} \mu M_3 
		\tan\beta \; C (m^2_{\sBot_1}, m^2_{\sBot_2}, M^2_3) \,,     \\
		\Delta_b^\text{higgsino} & = \frac{y_t^2}{16\pi^2} \mu A_t   
		\tan\beta \; C ( m^2_{\sTop_1}, m^2_{\sTop_2}, \mu^2 ) \,,
	\end{align}
\end{subequations}
where $m_{\sTop_{1,2}}$, \(m_{\tilde b_{1,2}}\) are the masses of the \sTop{}
and \sBot{} mass eigenstates, \(M_3\) denotes the gluino mass, and \(\mu\) is the
higgsino mass parameter. The function \(C(x, y, z)\) is given by
\begin{equation}\label{eq:C0}
	C( x, y, z ) = \frac{ x y \ln \frac{y}{x} + y z \ln \frac{z}{y}
		+ x z \ln \frac{x}{z}}{(x - y)(y - z)(x - z)} \;.
\end{equation}

These $\Delta_b$ corrections lead to an enhancement of $y_b$ especially for
large $\mu < 0$ and $A_t,M_3>0$  as both contributions are negative in this case
and reduce the denominator of \cref{eq:botres}. For $\Delta_b \to - 1$ the
bottom Yukawa coupling gets pushed into the non-perturbative regime. Taking
\cref{eq:botres} into account can lead to important effects of \sBot{} vevs in
the MSSM, see~\cite{Hollik:2015pra}, as will be visible in our numerical
analysis below. We also take into account a similar but numerically smaller
effect for the Yukawa coupling of the $\tau$ lepton,
$y_\tau$~\cite{Guasch:2001wv}.

\section{Constraints from Vacuum Stability in MSSM Benchmark Scenarios}
\label{sec:bench}
In the following we are going to present an example application of our method
for obtaining vacuum stability constraints. We will study vacuum stability
constraints for some of the MSSM benchmark scenarios defined in~\cite{Bahl2018a}
to illustrate the impact of the constraints and compare our method to previous
approaches.

\subsection{Vacuum Stability in the $M_h^{125}$ Scenario}\label{sec:standard}
\begin{figure}
	\centering
	\includegraphics[width=\textwidth]{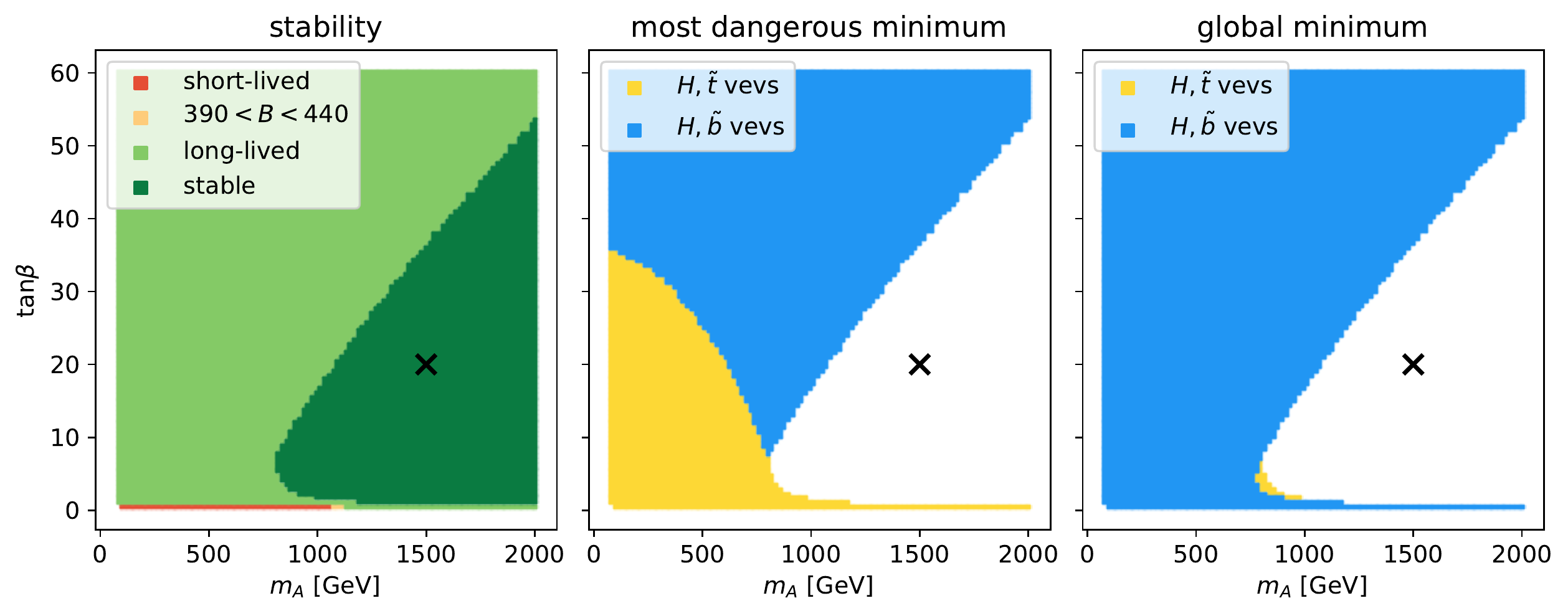}
	\caption{Constraints from vacuum stability in the $M_h^{125}$
		scenario defined in~\cite{Bahl2018a}. The colour code in the
		left panel indicates the lifetime of the EW vacuum, while the
		centre and right panels illustrate which fields have non-zero
		vevs at the most dangerous and the global minimum,
		respectively. The black $\boldsymbol{\times}$ marks the same
		point shown in \cref{fig:Mh125tri}.}\label{fig:Mh125}
\end{figure}

The first benchmark scenario defined in~\cite{Bahl2018a} is the
$M_h^{125}$ scenario. It features rather heavy SUSY particles with a SM-like
Higgs boson at $125\,\GeV{}$ and can be used
to display the sensitivity of searches for
additional Higgs bosons at the LHC\@.
Its parameters are given by
\begin{equation}
	\begin{gathered}
		m_{Q_3}=m_{U_3}=m_{D_3}=1.5\,\text{TeV}\,,\quad
		m_{L_3}=m_{E_3}=2\,\text{TeV}\,,\quad
		\mu=1\,\text{TeV}\,,\\
		X_t = A_t - \frac{\mu}{\tan\beta}=2.8\,\text{TeV}\,,\quad
		A_b=A_\tau=A_t\,,\\
		M_1=M_2=1\,\TeV\,,\quad M_3=2.5\,\TeV \, ,
	\end{gathered}\label{eq:parameters_std}
\end{equation}
while $m_A$ and $\tan\beta$ are varied in order to span the considered
parameter plane. The soft SUSY breaking parameters $A_{t,b,\tau}$ vary
as a function of $\tan\beta$ for fixed \(X_t\). Note that the gaugino
mass parameters $M_{1,2,3}$ only enter our analysis through the
$\Delta_b$ and $\Delta_\tau$ corrections.  \Cref{fig:Mh125} shows the
vacuum stability analysis in this benchmark plane.

In the left panel of \cref{fig:Mh125}, the colour code indicates the
lifetime of the EW vacuum at each point in the parameter plane. In the
\stablec{} region the EW vacuum is the global minimum of the theory, and
the EW vacuum is stable in this parameter region.  The \longlc{} area
depicts regions where deeper minima exist but the lifetime of the false
EW vacuum is large compared to the age of the universe (see
\cref{sec:lifetime}). For these parameter points the EW vacuum is
metastable and the parameter points are allowed. For points in the
\shortlc{} region, on the other hand, the tunnelling process is fast,
and they are excluded as the EW vacuum is short-lived. The small
\medlc{} region contains all points in the intermediate region discussed
in \cref{sec:lifetime} where an estimate of the uncertainties
gives no decisive conclusion on the longevity of the false vacuum. This
plot of \cref{fig:Mh125} left shows that the \(M_h^{125}\) benchmark
plane is hardly constrained by the requirement of vacuum stability.
Only a parameter region with small values of $\tan\beta\lesssim1$ can be
excluded.

The middle panel of \cref{fig:Mh125} shows the character of the most dangerous
minimum (MDM), \IE\ it displays which fields acquire non-vanishing vevs at this
vacuum. The MDM is defined as the minimum with the lowest bounce action for
tunnelling from the EW vacuum. One can see that for small values of $\tan\beta$
or both moderate $\tan\beta$ and $m_A$ the MDM is a CCB minimum with \sTop{}
vevs (\sTopvc{}). For larger values of $\tan\beta$ and $m_A$ a minimum with
\sBot{} vevs takes over (\sBotvc{}). This behaviour is expected as for higher
$\tan\beta$ the couplings of the Higgs sector to $d$-type \mbox{(s)quarks} are
enhanced which also increases the impact of \sBot{} vevs on our vacuum stability
analysis.

The right panel of \cref{fig:Mh125} displays the character of the global
minimum for the $M_h^{125}$ benchmark plane. It is important to note
that the MDM and the global minimum of the scalar potential
can in general differ from each other.\footnote{Out of the minima that are deeper than
	the EW vacuum, the MDM is usually the one that is closest to the EW
	vacuum in field space. However, this is not always the case, see
	\cref{fig:Mh125tri_vevcomp} below for a counterexample.} We see that
the global minimum has only \sBot{} vevs for most of the parameter
space, while there is a large region where the MDM involves
\sTop{} vevs.

\subsubsection{Impact of the Trilinear Terms}
\begin{figure}
	\centering
	\includegraphics[width=\textwidth]{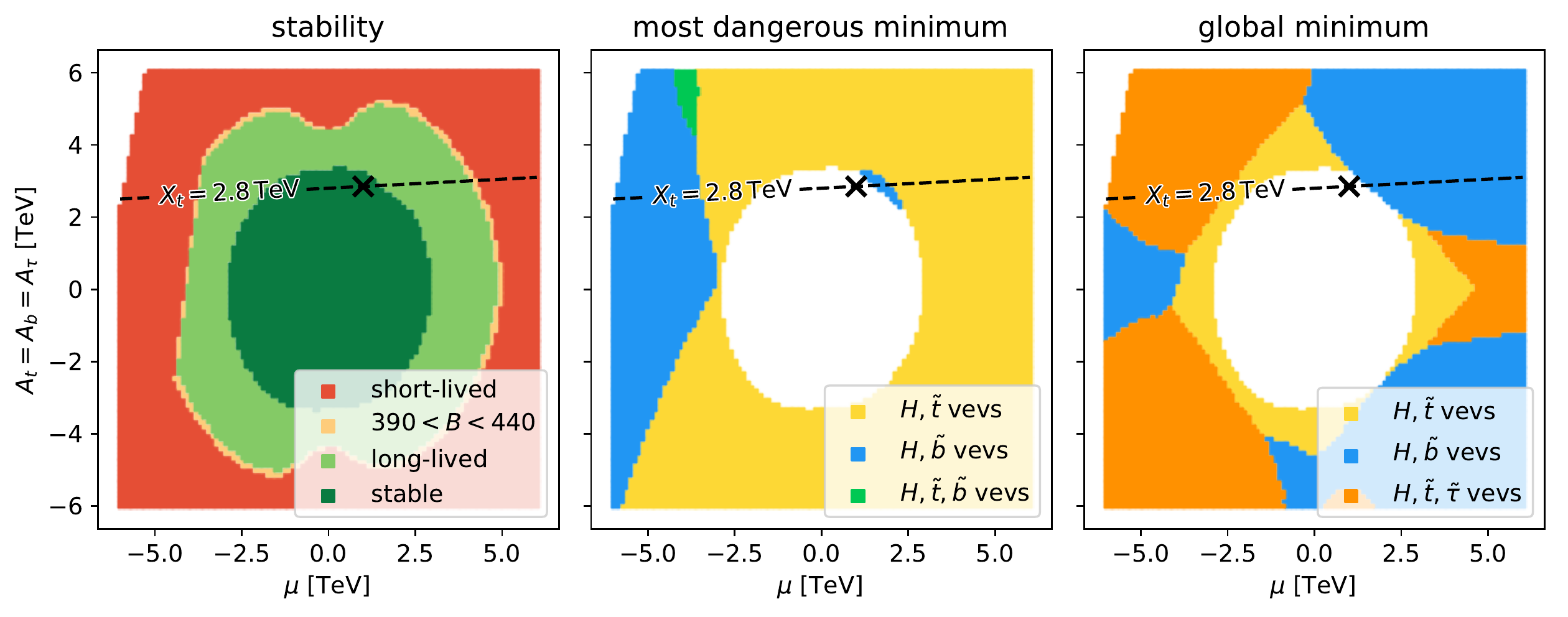}
	\caption{Constraints from vacuum stability in the plane of $\mu$ and $A$
		containing the selected point from the $M_h^{125}$ benchmark scenario. The
		starting point in the $M_h^{125}$ plane of \cref{fig:Mh125} with
		$\tan\beta=20$ and $m_A=1500$ is indicated by the black
		$\boldsymbol{\times}$. The colour code is the same as in \cref{fig:Mh125}.
		The dashed line corresponds to constant
		$X_t=2.8\,\TeV$.}\label{fig:Mh125tri}
\end{figure}

We noted in \cref{sec:deepdir} that the parameters entering the cubic terms of
the potential are expected to be especially important for the stability of the
EW vacuum. Since $m_A$ is related to a bilinear term in the potential, and
$\tan\beta$ mostly affects the quartic Yukawa couplings, we switch to a
different slice of the parameter space which is more relevant for the stability
studies. We start from a point in the $m_A$--$\tan\beta$ plane of the
\(M^{125}_h\) scenario that is absolutely stable and given by
\begin{equation}
	\tan\beta = 20\,,\quad m_A=1500\,\GeV\,.
\end{equation}
This point is indicated with a $\boldsymbol{\times}$ in
\cref{fig:Mh125,fig:Mh125tri}. It features a Higgs mass of
$m_h\approx125\,\GeV$ and is allowed by all the constraints considered
in~\cite{Bahl2018a}. Among these, the non-observation of heavy Higgs
bosons decaying into \(\tau\) pairs~\cite{Aaboud:2017sjh,
Sirunyan:2018zut} is the most relevant constraint.  In contrast to the
$m_A$--$\tan\beta$ plane of the \(M^{125}_h\) scenario, we now vary the
parameters $\mu$ and $A\equiv A_t=A_b=A_\tau$ starting from this point.
\Cref{fig:Mh125tri} shows the vacuum stability analysis in this new
parameter plane.

The left panel of \cref{fig:Mh125tri} indicates the lifetime of the EW
vacuum.  The colour coding is the same as in the corresponding plot of
\cref{fig:Mh125}, \IE\ \shortlc{} points depict short-lived
configurations, while the EW vacuum for \longlc{} points is
metastable, and the EW vacuum in the \stablec{} area is stable. The
thin \medlc{} band indicates the uncertainty band of $390 < B < 440$
discussed in \cref{sec:lifetime}. The EW vacuum becomes more and more
unstable for larger absolute values of $\mu$ and $A$. For small values
of these parameters the potential is absolutely stable with a region
of long-lived metastability in between. Note that also in this
parameter plane the \medlc{} uncertainty region of $390 < B < 440$
corresponds to only a thin band between long- and short-lived
regions. The point marked by the $\boldsymbol{\times}$ is the starting
point in the $M_h^{125}$ plane depicted in \cref{fig:Mh125}.
One can see that in the plane of \cref{fig:Mh125tri} this
point is close to a region of metastability, but quite far from any
dangerously short lived parameter regions. The missing points in the
top-left corner of the plot are points with tachyonic tree-level
\sBot{} masses where the EW vacuum is a saddle point.

The character of the MDM, \IE{}\ the fields that acquire non-zero vevs
in this vacuum, is shown in the middle panel of
\cref{fig:Mh125tri}. It is dominated by \sTopvc{} \sTop{} vevs in this
plane, but \sBotvc{} \sBot{} vevs are also important for large
negative values of $\mu$. The $\Delta_b$ corrections described in
\cref{sec:deltab} are enhanced in this parameter region and have a
large impact. They are also the cause of the tachyonic region for
large negative $\mu$ and positive $A$. Between the \sTop{}-vev and
\sBot{}-vev regime a region appears (shown in \sTopsBotvc{}) where
\sTop{} and \sBot{} vevs occur simultaneously. The
small \sBotvc{} region for $\mu>0$ is visible because the more
dangerous minima with \sTop{} vevs only appear for slightly higher
values of $A$ and $\mu$, and the global \sBot{}-vev
minimum is the only other vacuum in this parameter region besides the
EW vacuum.

In the right panel of \cref{fig:Mh125tri}, the fields which acquire non-zero
vevs at the global minimum are indicated. In this parameter plane, there are
large regions with simultaneous \sTop{} and \sTau{} vevs at the global minimum.
Through most of the plane the fields acquiring vevs differ between the MDM and
the global minimum. The \sTopsBotvc{} region of simultaneous \sTop{} and \sBot{}
vevs which is visible in the middle panel of \cref{fig:Mh125tri} does not
correspond to the global minimum of the theory. This is expected as additional
large quartic $F$ and $D$-term contributions appear if multiple kinds of squarks
take on non-zero vevs simultaneously. These are positive contributions to the
scalar potential that lift up these regions of field space.  No such
contributions appear in the case of simultaneous squark and slepton vevs which
is why the \sTopsTauvc{} regions of simultaneous \sTop{} and \sTau{} vevs are
present in the right panel of \cref{fig:Mh125tri}. Note that the quartic $F$ and
$D$-term contributions do not prevent the minima with mixed \sTop{} and \sBot{}
vevs from being the MDM as \cref{fig:Mh125tri} (centre) shows. However, for the
parameter plane considered here these minima featuring simultaneous \sTop{} and
\sTau{} vevs have no impact on the stability constraints of \cref{fig:Mh125tri}
(left).

We finally comment on the impact of the detailed field content, in particular
the first and second generation sfermions and the $\tilde\nu$ vevs, on
these results. The small Yukawa couplings of these particles tend to
push any additional minima to very large field values, which renders
these configurations long-lived. As a consequence, the metastability
bound (\medlc{} region in \cref{fig:Mh125tri}, left) and the character
of the MDM (\cref{fig:Mh125tri}, centre) are insensitive to the impact
of those fields. On the other hand, the character of the global minimum
may be significantly affected by fields with relatively small couplings
to the Higgs sector. Indeed, if we were to include $\tilde{c}$ and
$\tilde{s}$ vevs they would dominate the global minimum
(\cref{fig:Mh125tri}, right) through most of the parameter plane. They
would even cut slightly into the edges of the stable \stablec{} region
turning it long-lived. Our analysis therefore shows that neither the
investigation of just the region of absolute stability nor of the
character of the global minimum yields reliable bounds from vacuum
stability. This is due to the fact that both these quantities can
sensitively depend on the considered field content, where even very
weakly coupled scalar degrees of freedom can have a significant impact.
Instead, the correct determination of the boundary between the
short-lived and the long-lived region crucially relies on the correct
identification of the MDM, which in general can be very different from
the global minimum. This boundary, and accordingly the constraint on the
parameter space from vacuum stability, is in fact governed by the fields
with the largest Yukawa couplings and therefore insensitive to effects
from particles with a small coupling to the Higgs sector.

\subsubsection{Comparison to Semi-Analytic Bounds and Existing Codes}

We now compare our results shown in \cref{fig:Mh125tri} with results from the
literature. An approximate bound for MSSM CCB instabilities including vacuum
tunnelling is given by~\cite{Casas:1995pd, Kusenko:1996jn}
\begin{equation}
	A_t^2+3\mu^2 < (m_{\sTop_R}^2 + m_{\sTop_L}^2)\cdot
	\begin{cases}
		3\quad   & \text{stable,}     \\
		7.5\quad & \text{long-lived.} 
	\end{cases}\label{eq:sanabounds}
\end{equation}
Furthermore, a ``heuristic'' bound of
\begin{equation}
	\frac{\text{max}(A_{\sTop,\sBot},\mu)}{\text{min}(m_{Q_3,U_3})}\lesssim 3\label{eq:heuristic}
\end{equation}
is sometimes used to judge whether a parameter point might be sufficiently
long-lived (see \EG{} the discussion in~\cite{Bechtle:2016kui}).

The public code \texttt{Vevacious}~\cite{Camargo-Molina2013, Camargo-Molina2014}
can calculate the lifetime of the EW vacuum in BSM models using the
tree-level or Coleman--Weinberg one-loop potential, optionally including
finite temperature effects, see \EG{}~\cite{Chattopadhyay:2014gfa}.

\begin{figure}
	\centering
	\includegraphics[width=0.4\textwidth]{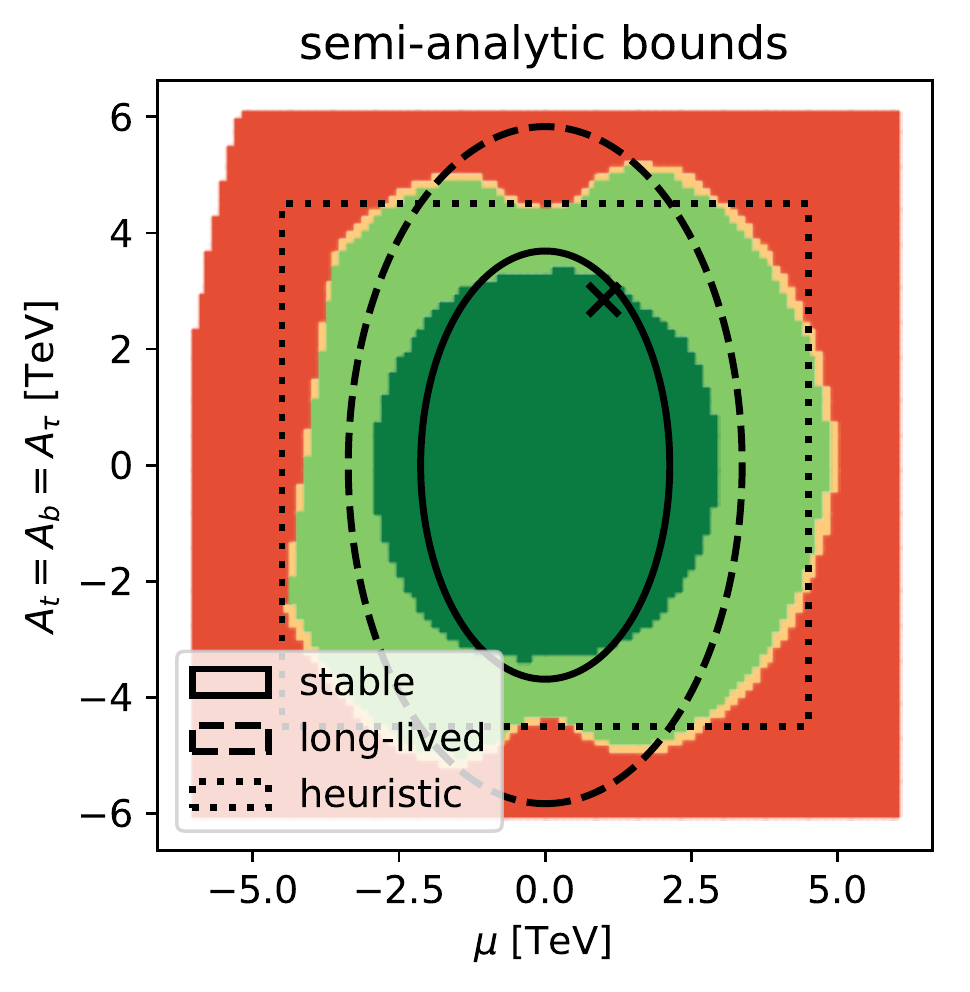}
	\caption{Constraints from vacuum stability in the plane of $\mu$ and $A$
		containing the selected point (black $\boldsymbol{\times}$) from the $M_h^{125}$ benchmark scenario. The results from \cref{fig:Mh125tri} are shown with
		superimposed contours indicating the approximate absolute and metastability
		bounds of \cref{eq:sanabounds} and the heuristic bound of
		\cref{eq:heuristic}.}\label{fig:Mh125tri_approxcomp}
\end{figure}

\Cref{fig:Mh125tri_approxcomp} displays our results in comparison to the
approximate bounds given in~\cref{eq:sanabounds,eq:heuristic}, where the
contours arising from~\cref{eq:sanabounds,eq:heuristic} are superimposed on our
results. The solid black contour arising from \cref{eq:sanabounds} should be
compared with the edge of the \stablec{} region where the vacuum is stable. This
comparison shows significant deviations. This is in particular due to the fact
that the absolute stability bound from \cref{eq:sanabounds} considers only the
$D$-flat direction. The dashed black contour should be compared with the
\medlc{} region at the border between the long-lived (\longlc{}) and short-lived
(\shortlc{}) regions and shows similar deviations. One source of these
deviations is that the metastability bound in \cref{eq:sanabounds} becomes less
reliable for values of $m_{\sTop_R}^2 + m_{\sTop_L}^2\gtrsim {(1200\,\GeV)}^2$ and
large \((A_{\tilde t}^2 + 3 \mu^2)\) (see Fig.~4 in~\cite{Kusenko:1996jn}).
Moreover, the dependence on $\tan\beta$~\cite{Blinov:2013fta} is not included in
the approximate bound. Another reason is that only \sTop{}-related parameters
enter \cref{eq:sanabounds}, while our analysis shows that also \sBot{} vevs have
important effects in this case.

The heuristic bound \cref{eq:heuristic} (dotted black contour in
\cref{fig:Mh125tri_approxcomp}) should also be compared to the \medlc{}
region.  While there are clear differences in shape, the size of the long-lived
region in our result roughly matches the heuristic bound. This can be
qualitatively understood from \cref{eq:deepdir}. For $\lambda\sim\mathcal{O}(1)$
this yields $A/m > 2$ as a bound for absolute stability. Therefore $A/m > 3$ as
a bound for metastability appears to be a reasonable estimate. While we only
show these comparisons for one parameter plane they hold very similarly for
every plane we have studied. Our comparison shows that all of these approximate
bounds have deficiencies in determining the allowed parameter region, and
dedicated analyses are necessary to obtain more reliable conclusions.

\begin{figure}
	\centering
	\includegraphics[width=\textwidth]{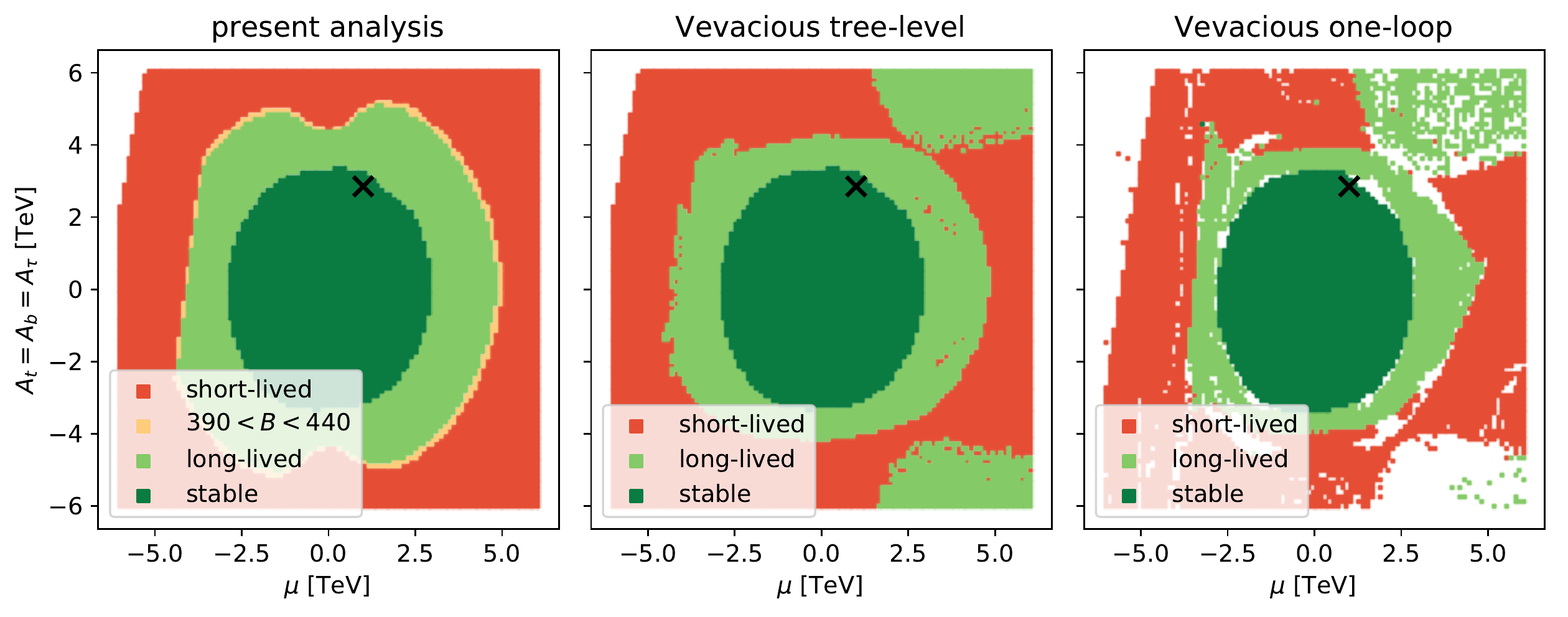}
	\caption{Constraints from vacuum stability in the plane of $\mu$ and $A$
		containing the selected point (black $\boldsymbol{\times}$) from the $M_h^{125}$ benchmark scenario. The results from \cref{fig:Mh125tri} are shown in the left
		panel. The other two plots show results of the code \texttt{Vevacious} for
		the tree-level (centre) and one-loop effective potential at zero temperature
		(right) for the same parameter plane.}\label{fig:Mh125tri_vevcomp}
\end{figure}

Next we compare our results (\cref{fig:Mh125tri_vevcomp}, left) to the
tree-level (\cref{fig:Mh125tri_vevcomp}, centre) and one loop (right) results of
\texttt{Vevacious}. In the \texttt{Vevacious} runs we have taken into account
only the fields from \cref{eq:fieldstb} since we found no relevant constraints
from \sTau{} vevs in this plane. \texttt{Vevacious} by default considers
tunnelling to the minimum which is closest in field space to the EW vacuum. In
the newest version (\texttt{1.2.03+}~\cite{vevaciousweb}) one can optionally
consider tunnelling to the global minimum instead. In generating
\cref{fig:Mh125tri_vevcomp} we combined the results from both of these
approaches by choosing the option giving the stronger bound at each individual
point. One obvious difference between our results and \texttt{Vevacious} are the
metastable regions that \texttt{Vevacious} finds for $\mu\sim3\,\TeV$ and
$|A|\sim5\,\TeV$. In this region \texttt{Vevacious} considers the wrong minimum
to be the MDM\@. The global minimum (with \sBot{} vevs, see \cref{fig:Mh125tri},
right) is closest to the EW vacuum in field space. Therefore, \texttt{Vevacious}
can only consider tunnelling into this minimum instead of a slightly further and
shallower minimum with \sTop{} vevs which gives the stronger constraints shown
in our results. {A similar issue is responsible for the edge in the
	\texttt{Vevacious} result around $\mu\sim-2.5\,\TeV$ and $A\sim4\,\TeV$.} A
second kind of visible difference is the absence in the \texttt{Vevacious}
result of the bumps in the long-lived region in our result around
$|\mu|\sim2\,\TeV$ and $|A|\sim 5\,\TeV$. The optimization of the bounce action
by \texttt{CosmoTransitions}~\cite{Wainwright:2011kj}, which is used by
\texttt{Vevacious}, leads to a slightly stronger and more reliable metastability
bound in this region.\footnote{As a cross check, forcing \texttt{Vevacious} to
use the direct path approximation yields the same lifetimes as our approach.}
Apart from these deviations our results are in good agreement with the
tree-level results of \texttt{Vevacious}. The deviations for individual points
and the rugged edges of the \longlc{} region in the \texttt{Vevacious} result
are likely signs of numerical instability. This especially includes the isolated
\shortlc{} points in the \longlc{} region which result from numerical errors in
the calculation of the tunnelling time.

The comparison with the \texttt{Vevacious} results using the Coleman--Weinberg
one-loop effective potential at zero temperature (\cref{fig:Mh125tri_vevcomp},
right) shows that the one-loop effects on the allowed parameter space are small
for this scenario. The one-loop result from \texttt{Vevacious} clearly suffers
from numerical instabilities. However, the stable region is nearly identical to
the tree-level results, and the long-lived region is similarly sized as the
tree-level \texttt{Vevacious} result with differences in shape. The long-lived
region appearing around $A\sim 5\,\TeV$ and $\mu\sim 4\,\TeV$ as well as the
missing region around $A\sim -5\,\TeV$ and the spikes around $\mu\sim-2.5\,\TeV$
are consequences of the same MDM misidentification as in the \texttt{Vevacious}
tree-level result (see previous paragraph). Comparing the runtime of our code to
the runtime of \texttt{Vevacious} in this parameter plane including only the
field set of \cref{eq:fieldstb} we find our tree-level code to be $\sim5$ times
faster than the tree-level and $\sim200$ times faster than the one-loop
\texttt{Vevacious} run.

\subsubsection{Parameter Dependence of the Vacuum Structure and
Degenerate Vacua}

The dashed line in \cref{fig:Mh125tri} is the line where $X_t$ has the
same value as in the benchmark plane, \cref{fig:Mh125}. The mass $m_h$
of the SM-like Higgs boson depends dominantly on the parameters
$\tan\beta$, $X_t$ and the stop masses. We therefore expect the Higgs
mass to stay close to $125\,\GeV$ when moving away from the point
$\boldsymbol{\times}$ along this line.\footnote{We have verified using
	\texttt{FeynHiggs 2.14.3}~\cite{Heinemeyer1999, Heinemeyer2000,
	Degrassi2003, Frank2007, Hahn2014, Bahl2016, Bahl2018} that
	$124\,\GeV \lesssim m_h \lesssim 126\,\GeV$ indeed holds along this
	line as long as $|\mu|\lesssim 3\,\TeV$.} We use this as motivation to
further investigate the vacuum structure along this line.

\begin{figure}
	\centering
	\includegraphics[width=0.8\textwidth]{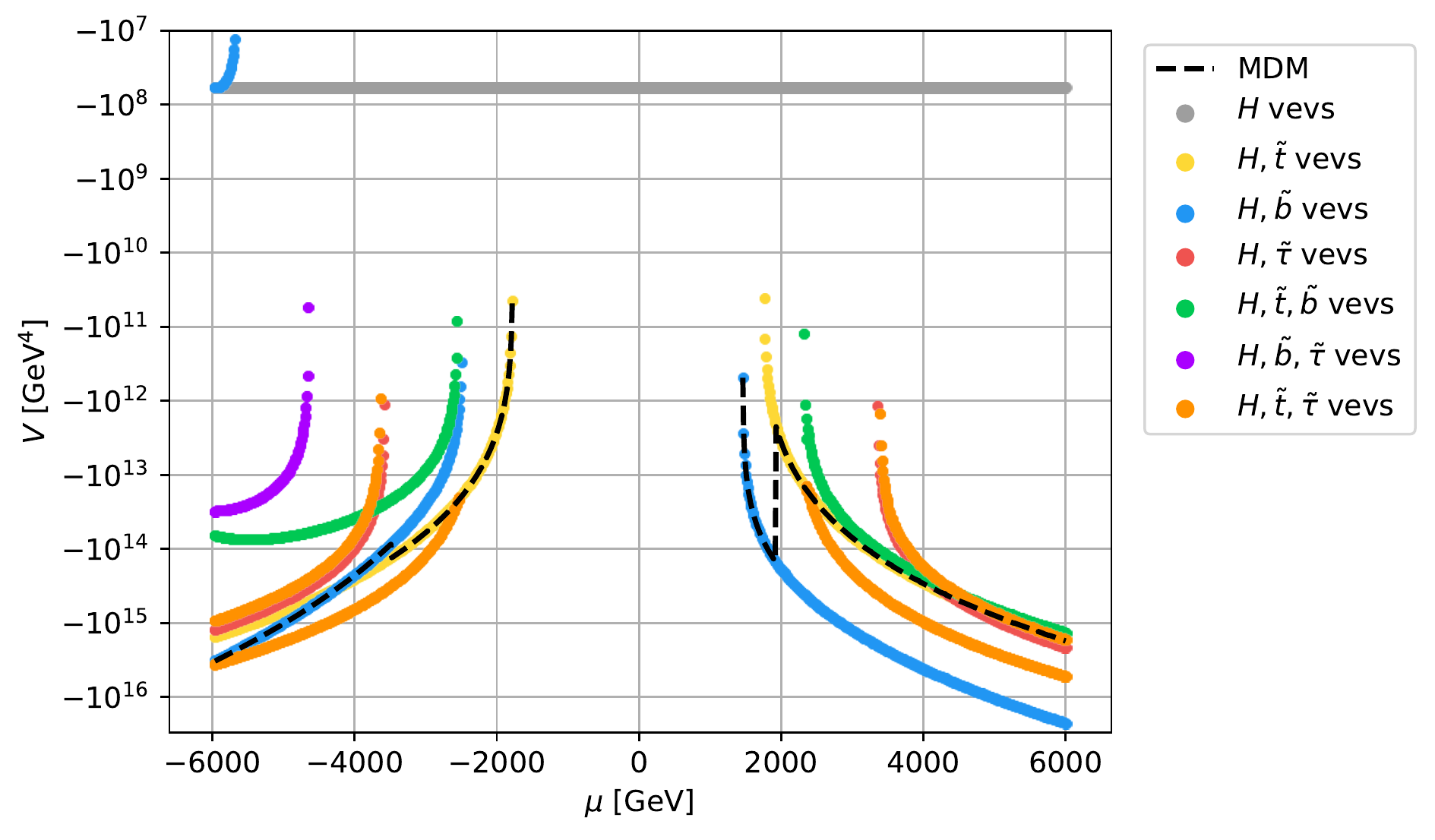}
	\caption{Depth of the different types of stationary points
		along the line of constant $X_t=2.8\,\TeV$ from
		\cref{fig:Mh125tri}. The colour code indicates which fields
		acquire vevs at the stationary point. The dashed line
		indicates which of the stationary points is the MDM\@. The
		grey line is the EW vacuum.}\label{fig:Mh125triline}
\end{figure}

\Cref{fig:Mh125triline} shows the depth of the stationary points of
the scalar potential as a function of $\mu$ along this line. The
constant depth of the EW vacuum is shown in grey while the other
colours indicate the CCB stationary points.
Note that
not only local minima, but all stationary points including saddle
points and local maxima are shown in \cref{fig:Mh125triline}.
The dashed line indicates the MDM for each value of $\mu$.

It can be seen from \cref{fig:Mh125triline} that for large negative \(\mu\)
simultaneous \sTop{}  and \sTau{} vevs (\sTopsTauvc{}) dominate the global
minimum for the considered field content until the \sTau{} vevs at these
stationary points approach zero around $\mu=-2.2\,\TeV$, and pure \sTop{} vevs
take over. From $\mu\approx -1.8\,\TeV$ onwards the EW vacuum is the global
minimum until a CCB vacuum with \sBot{} vevs appears at $\mu\approx 1.6\,\TeV$.
The MDM, on the other hand, is the second deepest \sBot{}-vev minimum for
$\mu\lesssim-3.5\,\TeV$, before switching to the \sTop{}-vev minimum, followed
by the window of absolute stability \(\mu \in [-1.8\,\TeV, 1.5\,\TeV]\). For
positive values of $\mu > 1.5\,\TeV$ the instability first develops towards the
global \sBot{}-vev minimum until the \sTop{}-vev minimum takes over at
$\mu\approx2\,\TeV$.

In \cref{fig:Mh125triline} several stationary points with multiple kinds of
sfermion vevs appear. Stationary points with mixed squark and slepton vevs can
be deeper than the corresponding stationary points with only one type of vev
(this can be seen for instance by comparing the deepest stationary point with
\sTopsTauvc{} \sTop{} and \sTau{} vevs to the ones with \sTopvc{} \sTop{} vevs
and \sTauvc{} \sTau{} vevs). A stationary point with multiple kinds of squark
vevs, however, is always higher than one with only one kind of the involved
vevs. This is due to the additional positive quartic contributions to the
potential for stationary points with both kinds of squark vevs.

Another feature visible in \cref{fig:Mh125triline} is the \sBot{}-vev stationary
point approaching the EW vacuum at $\mu\ll0$ from above. In this regime the
$\Delta_b$ corrections significantly enhance the bottom Yukawa coupling giving
rise to a large mixing in the \sBot{} sector and a corresponding decrease of one
of the \sBot{} masses. The depth of this stationary point becomes degenerate
with the EW vacuum. For even larger negative \(\mu\) one \sBot{} squark becomes
tachyonic, and the EW vacuum turns into a saddle point. The plot ends before
this happens (corresponding to the white region in \cref{fig:Mh125tri}) as we
require the existence of an EW vacuum.

The scalar potential \cref{eq:MSSMpot} for our field sets
\cref{eq:fieldstb,eq:fieldstl,eq:fieldsbl} has two accidental
$\mathbb{Z}_2$ symmetries. The potential is symmetric under
simultaneous sign flips of the left- and right-handed sfermions of a
kind
\begin{equation}
	\Re(\tilde{f}_L),\,\Re(\tilde{f}_R)\rightarrow
	-\Re(\tilde{f}_L),\,-\Re(\tilde{f}_R)
	\quad\text{with } \tilde{f} \in \{\sTop,\sBot,\sTau\}\label{eq:minsymsquark}
\end{equation}
and under simultaneous sign flips of all doublet components
\begin{equation}
	\begin{aligned}
		                 & \Re(h_u^{0}),\,\Re(h_d^{0}),\,\Re(\sTop_L),\,\Re(\sBot_L),\,\Re(\sTau_L)                           \\
		\rightarrow\quad & -\Re(h_u^{0}),\,-\Re(h_d^{0}),\,-\Re(\sTop_L),\,-\Re(\sBot_L),\,-\Re(\sTau_L)\,.\label{eq:minsymh} 
	\end{aligned}
\end{equation}
This results in sets of degenerate and physically equivalent stationary
points related by these symmetries.\footnote{Since these minima are
	degenerate they cannot be distinguished in \cref{fig:Mh125triline}.}
Since the EW vacuum is also invariant under \cref{eq:minsymsquark} the
tunnelling time to minima related by this symmetry is always
identical. However, since the EW vacuum breaks
\cref{eq:minsymh}\footnote{We can, without loss of generality, choose
	the EW vacuum with $\Re(h_u^{0}),\,\Re(h_d^{0}) > 0$.} the tunnelling
time into two stationary points related through this transformation can
differ. In most cases, whichever of these two points is closer in
field space to the EW vacuum gives the lower value for $B$. Note that
this is not a small effect. The values of $B$ for stationary points
related by \cref{eq:minsymh} can differ by more than an order of
magnitude. This effect has recently been studied for the simpler case of
a 2HDM in~\cite{Branchina2018}.

\subsection{Vacuum Stability in the $M_h^{125}(\sTau{})$ Scenario}

A benchmark scenario with light \sTau{} has been proposed in~\cite{Bahl2018a} under
the name $M_h^{125}(\sTau{})$. It is defined by
\begin{equation}
	\begin{gathered}
		m_{Q_3}=m_{U_3}=m_{D_3}=1.5\,\TeV\,,\quad m_{L_3}=m_{E_3}=350\,\GeV\,,\quad
		\mu=1\,\TeV\,,\\
		X_t = A_t - \frac{\mu}{\tan\beta}=2.8\,\TeV\,,
		\quad A_b=A_t\,,\quad A_\tau=800\,\GeV\,,\\
		M_1=M_2=1\,\TeV\,,\quad M_3=2.5\,\TeV\,.
	\end{gathered}
\end{equation}
The scenario differs from the $M_h^{125}$ scenario of \cref{eq:parameters_std}
only in greatly reduced soft \sTau{} masses with a correspondingly reduced
$A_{\sTau{}}$. However, $\mu$ is not reduced and is now $\mu\sim 3 m_{\sTau{}}$.
According to \cref{eq:heuristic} we would therefore expect vacuum stability
constraints to be relevant in the $M_h^{125}(\sTau{})$ benchmark plane. The
authors of~\cite{Bahl2018a} used \texttt{Vevacious} to check for vacuum
instabilities in this scenario and found a short-lived region in the parameter
space for large $\tan\beta$ and small $m_A$.

\begin{figure}
	\centering
	\includegraphics[width=\textwidth]{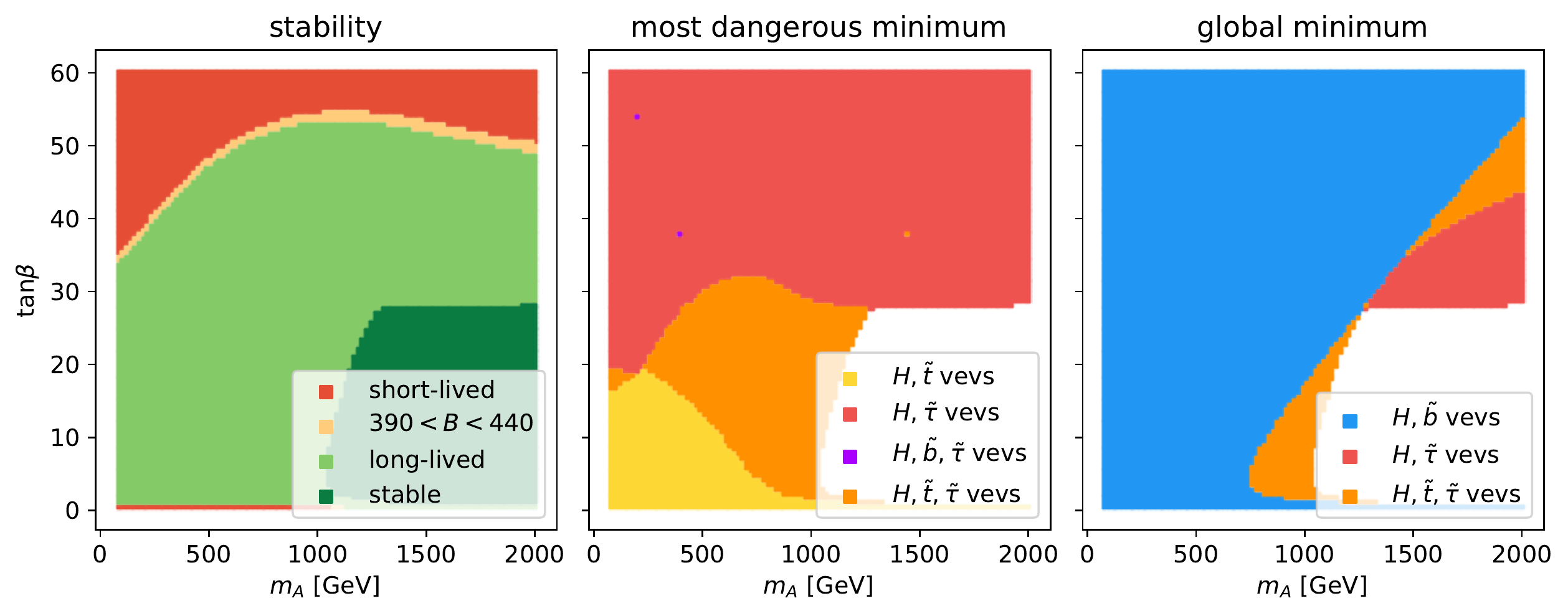}
	\caption{Constraints from vacuum stability in the
		$M_h^{125}(\sTau)$ scenario.  The colour code in the left plot
		indicates the lifetime of the EW vacuum, while the centre and
		right plots illustrate which fields have non-zero vevs at the
		MDM and the global minimum,
		respectively.}\label{fig:Mh125stau}
\end{figure}

Our results shown in \cref{fig:Mh125stau} confirm these observations.
\Cref{fig:Mh125stau} (left) shows a short-lived region for large $\tan\beta$.
This region extends towards smaller values of $\tan\beta$ in the low $m_A$
regime as noted in~\cite{Bahl2018a} but also in the region of large $m_A$. Also
visible is the small region of instability for $\tan\beta < 1$ noted in
\cref{fig:Mh125}. The MDM for the instability at large $\tan\beta$ is a vacuum
with \sTau{} vevs as can bee seen from \cref{fig:Mh125stau} (centre). Compared
to \cref{fig:Mh125} the absolutely stable region is additionally reduced by a
\sTop{}-\sTau{}-vev minimum appearing around $m_A\sim1\,\TeV$ and
$\tan\beta<30$. The minima with \sBot{} vevs, which were the MDM for large
regions of the $M_h^{125}$ scenario, are entirely replaced by minima with
\sTau{} vevs. Only a very small \sBotsTauvc{} region with simultaneous \sBot{}
and \sTau{} vevs at the MDM exists. In \cref{fig:Mh125stau} (right) the global
minimum with \sBot{} vevs is very similar to \cref{fig:Mh125} (right). Only at
larger $m_A$ --- where the EW vacuum in the $M_h^{125}$ scenario was absolutely
stable --- global minima with \sTau{} vevs are now present. Our results in the
$M_h^{125}(\sTau)$ scenario compared to the $M_h^{125}$ scenario illustrate that
constraints from vacuum stability indeed become relevant when the cubic terms in
the scalar potential become larger than the quadratic terms. Since
$\mu\sim3m_{\sTau{}}$ in this scenario, a further increase of $\mu$ or a
decrease of $m_{\sTau{}}$ could render the $M_h^{125}(\sTau)$ scenario entirely
short-lived. In the region of $\mu\sim3m_{\sTau{}}$ chosen in the
$M_h^{125}(\sTau)$ scenario, the vacuum stability constraints show a
significant dependence on the parameters $m_A$ and $\tan\beta$.

\subsection{Vacuum Stability in the $M_h^{125}(\mathrm{alignment})$
	Scenario}\label{sec:ali}

In this section we turn to another scenario from~\cite{Bahl2018a}, the
$M_h^{125}(\text{alignment})$ scenario, where constraints from vacuum
stability turn out to have a very large impact. The scenario is defined
by
\begin{equation}
	\begin{gathered}
		m_{Q_3}=m_{U_3}=m_{D_3}=2.5\,\TeV\,,\quad m_{L_3}=m_{E_3}=2\,\TeV\,,\\
		\mu=7.5\,\TeV\,,\quad A_t=A_b=A_\tau=6.25\,\TeV\,,\\
		M_1 = 500\,\GeV\,,\quad M_2 = 1\,\TeV\,,\quad M_3=2.5\,\TeV\,,
	\end{gathered}\label{eq:parameters_ali}
\end{equation}
with $m_A\in[100\,\GeV,1\,\TeV]$ and $\tan\beta\in[1,20]$. It is chosen to
accommodate light additional Higgs bosons through the so-called alignment
without decoupling behaviour~\cite{Gunion2003, Craig2013,Carena2014, Dev:2014yca, 
Carena2015, Bernon2015, Bernon2016, Bechtle:2016kui, Haber2017}. Alignment
without decoupling in the phenomenologically interesting region of relatively
small $\tan\beta$ requires $\mu,A_t\gg m_{\sTop}$. This requirement---according
to \cref{eq:heuristic}---has already been noted to be problematic for vacuum
stability in~\cite{Bahl2018a}. Indeed we find that the EW vacuum in this
scenario is short-lived through all of its parameter space with instabilities in
directions with \sTop{} vevs. The corresponding \cref{fig:Mh125ali} can be found
in \cref{app:rest}.

\begin{figure}
	\centering
	\includegraphics[width=\textwidth]{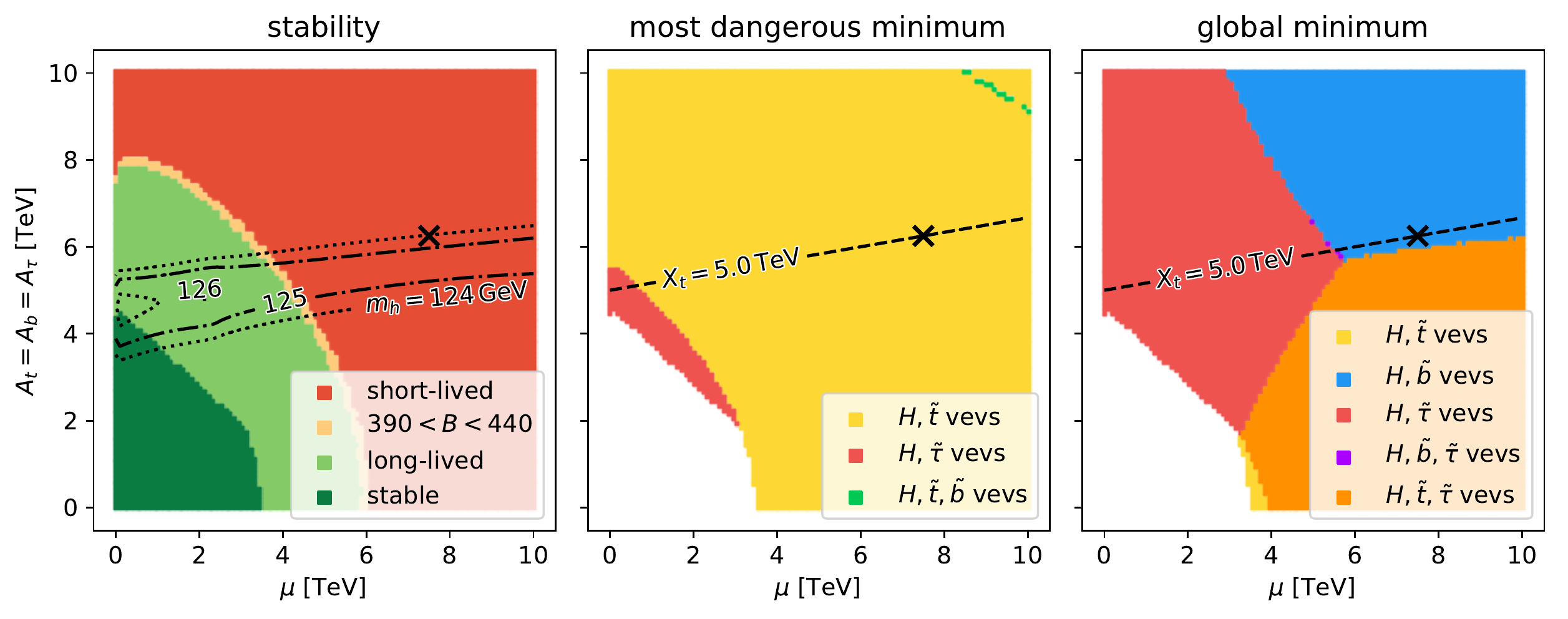}
	\caption{Constraints from vacuum stability in the plane of $\mu$ and $A$
		containing the selected point from the $M_h^{125}(\text{alignment})$
		scenario marked with the black $\boldsymbol{\times}$. The
		selected point in the \(m_A\)-\(\tan\beta\) plane of the
		$M_h^{125}(\text{alignment})$ scenario with
		$\tan\beta=6$ and $m_A=500\,\GeV$ is
		shown in \cref{fig:Mh125ali}, see \cref{app:rest}. The colour code in the
		left plot indicates the lifetime of the EW vacuum, while the centre and
		right plots illustrate which fields have non-zero vevs at the MDM and the
		global minimum, respectively.  The contours in the left plot indicate the
		mass of the Higgs boson $h$, and the dashed line shown in
		the centre and right panels indicates the line of constant
		$X_t=5\,\TeV$.}\label{fig:Mh125alitri}
\end{figure}

In order to assess how far away this scenario is from a region of
metastability we again select a phenomenologically interesting point in
its parameter plane,
\begin{equation}
	m_A=500\,\GeV\,,\quad \tan\beta=6\,,
\end{equation}
and vary $A=A_t=A_b=A_\tau$ and $\mu$ starting from this point. The
resulting plane is shown in \cref{fig:Mh125alitri}. The colour code and
the quantities shown in the sub-plots are the same as in
\cref{fig:Mh125,fig:Mh125tri}. In the left panel we have superimposed
the contours for the mass of the light Higgs boson calculated with
\texttt{FeynHiggs 2.14.3}~\cite{Heinemeyer1999, Heinemeyer2000,
Degrassi2003, Frank2007, Hahn2014, Bahl2016, Bahl2018}.  This shows
that in order to obtain a long-lived scenario while keeping the correct
Higgs mass one could for example change
\begin{equation}
	\mu = 7.5\,\TeV\to 4\,\TeV
	\quad\text{and}\quad
	A = 6.25\,\TeV\to 5\,\TeV\,.\label{eq:shiftedali}
\end{equation}
Other choices are possible within the theoretical uncertainty of the Higgs mass
prediction. The alignment-without-decoupling behaviour implies that the
$125\,\GeV$ Higgs boson has SM-like properties. In the MSSM this behaviour
arises from a cancellation between tree-level mixing contributions and
higher-order corrections in the Higgs-boson mass matrix. Shifting the parameters
towards the region where the vacuum is metastable like it is done in
\cref{eq:shiftedali} could affect these cancellations and therefore spoil the
alignment properties. Indeed, with \texttt{FeynHiggs} we obtain $\sim 10\%$
enhanced couplings of $h$ to bottom and tau pairs leading to a $\sim 10\%$
reduced $\text{BR}(h\to\gamma\gamma)$ compared to the SM for the parameter point
of \cref{eq:shiftedali}. While the change in the couplings indicates that the
cancellation that is present in the alignment without decoupling regime does not
fully occur for the shifted point, this parameter point is nevertheless still
compatible with the LHC Higgs measurements within the present
uncertainties~\cite{Aaboud:2018xdt}.

\begin{figure}
	\centering
	\includegraphics[width=0.8\textwidth]{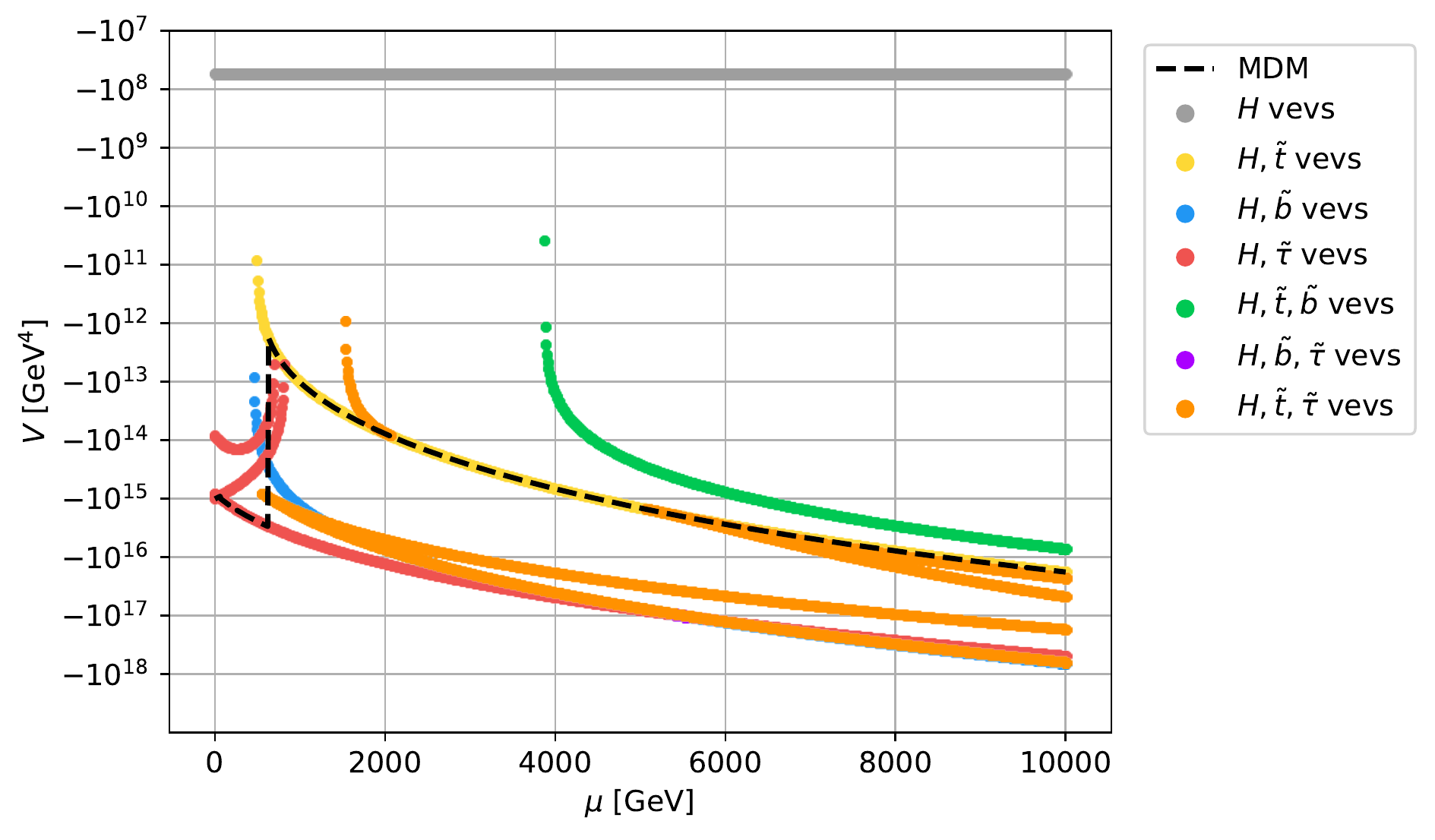}
	\caption{Depth of the different types of stationary points along
		the line of constant $X_t=5\,\TeV$ from
		\cref{fig:Mh125alitri}. The colour code indicates which fields
		have non-zero vevs at the stationary point. The dashed line indicates
		which of the stationary points is the MDM\@. The EW vacuum is
		shown in gray.}\label{fig:Mh125alitriline}
\end{figure}

In the middle panel of \cref{fig:Mh125alitri} one can see that the MDM
directions through most of the parameter space are directions with \sTop{} vevs.
If the instabilities were in directions with \sBot{} or \sTau{} vevs, the
requirement of a long-lived vacuum could have been achieved by increasing
$m_{D_3,L_3,E_3}$ and/or decreasing $A_{b,\tau}$ while keeping $\mu$ and the
\sTop{} sector unchanged such that the phenomenology would not be much affected.
However, our analysis shows that the parameter region associated with the
alignment-without-decoupling behaviour gives rise to minima with \sTop{} vevs
that render the electroweak vacuum short-lived. Accordingly, the behaviour of
alignment without decoupling and the requirement of a long-lived vacuum are in
some tension with each other, since adjusting $m_{Q_3,U_3}$ and $A_t$ to ensure
vacuum stability in the stop directions would change the mass and phenomenology
of the light Higgs boson $h$. The right plot of \cref{fig:Mh125alitri}
illustrates that the global minimum has non-vanishing \sTau{} vevs through
significant parts of the parameter plane, while for large values of \(A, \mu
\gtrsim 5\,\TeV\) \sBot{} vevs take over. However, the deepest minima appear to
be nearly degenerate as can be seen from \cref{fig:Mh125alitriline}.

The depth of the stationary points of the scalar potential along the line of
constant $X_t=5\,\TeV$ (which is indicated in \cref{fig:Mh125alitri} centre and
right) is shown in \cref{fig:Mh125alitriline}. It includes all stationary points
in the selected field sets that are at least as deep as the EW vacuum. The EW
vacuum is shown in grey, and the CCB stationary points are distinguished by the
other colours. The plot illustrates that there is no stable region along this
slice of parameter space, in accordance with \cref{fig:Mh125alitri}. As already
pointed out, the MDM along this line is a minimum with \sTop{} vevs through most
of the parameter range, with the exception of the region with small values of
$\mu$ where the MDM has \sTau{} vevs. There exist stationary points that are
deeper than the MDM with \sBot{} vevs\footnote{The \sBot{}-vev stationary point
	is almost degenerate with the deepest \sTop{}-\sTau-vev stationary point for
	most of the parameter range and therefore hidden behind the \sTopsTauvc{}
	points.}, \sTop{}-\sTau{} vevs and \sTau{} vevs with the \sTau{}-vev minimum
being the global minimum until the \sBot{}-vev minimum takes over for
$\mu\gtrsim 5.5\,\TeV$. These deeper minima are however very far from the EW
vacuum in field space with high barriers and have no impact on the tunnelling
rate. For large values of $\mu$ stationary points with both \sTop{} and \sBot{}
vevs develop. In agreement with the previous discussions these stationary points
are less deep than the stationary points with only one kind of squark vev. All
points along this line of constant $X_t=5\,\TeV$ would be long-lived if the MDM
with \sTop{} vevs was absent.

A detailed analysis of the question whether adjustments of the proposed
scenarios for alignment without decoupling could bring these scenarios into
better agreement with the constraints from vacuum stability while retaining
their alignment properties goes beyond the scope of this work. We leave this
issue for future studies.

\section{Summary and Conclusions}
\label{sec:conclusion}
In this work we have presented a fast and efficient method for determining the
constraints from vacuum stability on the model parameters of multi-scalar
theories beyond the Standard Model. The stability of the EW vacuum at the tree
level is investigated using polynomial homotopy continuation for the
minimization of the scalar potential and approximating the decay rate of the EW
vacuum into a deeper vacuum by an exact solution of the bounce action in the
one-field case. The method has been designed to combine a fast evaluation with
high numerical stability, enabling a reliable use in large scale parameter
scans. The generic approach admits a wide range of applications in many
different BSM models. We have argued that this approach is appropriate for
applications in multi-dimensional parameter spaces since the dependence of the
vacuum stability constraint on the model parameters is typically much stronger
than the impact of high precision calculations of the bounce path and the
tunnelling action. Therefore, the constraints on the parameter space are
relatively insensitive on the details of the calculation of the lifetime, and
uncertainties in the classification of long- and short-lived configurations only
affect small regions in the parameter space.

In order to evaluate the constraints from vacuum stability it is in principle
desirable to simultaneously take into account all possible vevs of all the
scalar fields in the considered model. However, this often becomes impractical
in models with large scalar sectors. We have illustrated our approach by
determining the constraints from vacuum stability for a set of benchmark
scenarios for Higgs searches in the MSSM that have recently been proposed
in~\cite{Bahl2018a}. Our efficient approach has allowed us to consider up to six
real degrees of freedom simultaneously. We have included the possible vevs of
the particles with the largest Yukawa couplings (namely \sTop{}, \sBot{} and
\sTau{}) in addition to the Higgs vevs when searching for unstable directions.
We have observed that this yields reliable vacuum stability
constraints---especially when including multiple kinds of sfermions
simultaneously. However, for the studied scenarios we would have obtained very
similar vacuum stability constraints if we had considered each kind of sfermion
separately as an approximation.

The result that the sfermions of the first and second generation and the
sneutrinos have a minor impact on the determination of the boundary between
long-lived and short-lived vacua is related to the fact that their smaller
Yukawa couplings lead to additional minima at very large field values, which
renders these configurations long-lived. On the other hand, the global
minimum---and to some extent also the region of absolute stability---shows a
significant dependence on additional field content that couples only weakly to
the Higgs sector. We also found several parameter regions where the global
minimum is characterised by vevs of different sfermions that are simultaneously
non-zero.

As a result, our analysis shows that neither the investigation of just
the region of absolute stability nor of the character of the global
minimum is sufficient to obtain reliable bounds from vacuum stability.
Instead, the determination of the boundary between the short-lived
and the long-lived region crucially relies on the correct
identification of the most dangerous minimum (MDM), which is
the minimum with the shortest tunnelling time from the EW vacuum.
The MDM often differs from the global minimum.

For the considered $M_h^{125}$ MSSM benchmark scenario we have found that the
impact of vacuum stability constraints is small in the $m_A$--$\tan\beta$
parameter plane defining the scenario. However, a variation of the parameters
$\mu$ and $A$ around their values chosen in the benchmark scenario shows an
important impact of vacuum stability constraints. In this plane we have
illustrated that the most dangerous and the global minimum are in general
different. We have furthermore stressed the importance of corrections to the
relation between the bottom-quark mass and the bottom Yukawa coupling in certain
regions of parameter space. These $\Delta_b$ corrections can significantly
enhance the value of the bottom Yukawa coupling and thus trigger \sBot{}-vev instabilities.

We have also used this parameter plane to compare our results with existing
studies and codes. Comparing our results to approximate analytic vacuum
stability bounds we have seen that those approximations can serve as a rough
estimate of the effect of vacuum stability constraints. However, they cannot
capture the complexity of a detailed numerical analysis. Furthermore, we have
compared the results of our code to the public code \texttt{Vevacious}. The
tree-level results of the codes show some notable differences. The largest
differences arise from the determination of the MDM\@. By default,
\texttt{Vevacious} uses the closest minimum in field space---and can be
optionally forced to use the global minimum. In contrast, by the use of an
analytic expression for the bounce action, we are able to easily calculate all
tunnelling times to deeper minima and select the one with the shortest time as
the MDM\@. We have seen that in the benchmark plane under consideration there
are regions where the MDM is neither the global minimum nor the minimum closest
to the EW vacuum in field space. We have also seen that the impact of the more
sophisticated tunnelling path calculation in \texttt{Vevacious} (using the path
deformation of \texttt{CosmoTransitions}) is visible but in most cases does not
substantially change the boundary between parameter regions with long- and
short-lived EW vacuum. Calculating the vacuum stability bounds in
\texttt{Vevacious} using the one-loop effective potential leads to results
qualitatively similar to the tree level results. The one-loop calculation showed
strong indications of numerical instability in addition to the problems in the
identification of the MDM\@. Even ignoring the latter problem we have
found the potential improvement by the one-loop effective potential on the
vacuum stability constraints to be relatively small in view of the significantly
increased amount of numerical instabilities and the much longer runtime.
Furthermore, as discussed above it is still an open conceptual question whether
the application of a loop-corrected effective potential can at all be expected
to yield a systematic improvement of vacuum stability constraints compared to a
tree-level analysis. The tree-level constraints of our approach using the
straight tunnelling path approximation yield numerically stable results in a
fraction of the runtime of \texttt{Vevacious} and \texttt{CosmoTransitions}.

We have exemplarily studied the depths of the different stationary points along
a line of $m_h\sim125\,\GeV$ through the parameter space. This has shown a
typical number of stationary points appearing in an analysis of vacuum stability
and illustrated the importance of the different scalar degrees of freedom. The
MDM was found to coincide with the global minimum only within a restricted part
of the relevant parameter range. We have also discussed the impact of degenerate
stationary points that are related to each other by a discrete symmetry and
found that---in agreement with recent results in the
literature~\cite{Branchina2018}---if such a symmetry is broken by the EW vacuum
the tunnelling times into the degenerate vacua can be vastly different.

In a benchmark scenario with light \sTau{} we have shown important constraints
from vacuum stability arising in the $m_A$--$\tan\beta$ plane defining the
scenario. In~\cite{Bahl2018a} a parameter region with high $\tan\beta$ and low
$m_A$ of this benchmark plane was identified as being excluded by vacuum
stability constraints. Our results provide a more detailed study of these
constraints and show that the excluded region extends to high $m_A$.

A further scenario that we have studied in detail is a benchmark scenario in the
alignment without decoupling regime. We have found that all parameter points in
the benchmark plane yield short-lived EW vacua. We have shown what kind of shift
in parameter space could lead to a sufficiently long-lived vacuum while
approximately preserving the correct value of the Higgs mass. We found that this
na\"ive approach would not obviously spoil the alignment without decouplings
behaviour. The question to what extent the alignment without decoupling
behaviour can be realised in phenomenologically viable scenarios where vacuum
stability constraints are taken into account would require a detailed study that
is beyond the scope of the present work. The remaining CP-conserving benchmark
scenarios\footnote{It should be noted that we have omitted the CP-violating
	scenario that was proposed in~\cite{Bahl2018a} for simplicity only. Our approach
is fully applicable also to models containing CP-violating phases.}
of~\cite{Bahl2018a} behave similarly to the ones we have studied. The vacuum
stability constraints in these benchmark planes are shown in \cref{app:rest}.

We plan to make our implementation of the described procedure publicly
available, which is meant to serve as an easily applicable tool for
evaluating vacuum stability constraints that is suitable for the inclusion
in scans over multi-dimensional parameter spaces.

\acknowledgments{
	We thank the authors of \texttt{Vevacious}, especially Jose Eliel Camargo-Molina
	and Werner Porod, for helpful comments and discussions.  We also thank Tim
	Stefaniak and Stefan Liebler for useful discussions. JW gratefully acknowledges
funding from the PIER Helmholtz Graduate School.}

\appendix

\section{The MSSM Scalar Potential}\label{app:pot}
In this appendix we give the scalar potential of the MSSM in expanded form. We
include all fields appearing in \cref{eq:fieldstb,eq:fieldstl,eq:fieldsbl}.
The potential is given by
\begin{equation}
	V = V_2+V_3+V_4\,,
\end{equation}
consisting of a quadratic ($V_2$), cubic ($V_3$) and quartic ($V_4$) part.
The quadratic part is given by
\begin{align}
	V_2 & =\begin{aligned}[t]                                                                                                                                    
	    & \frac{m_{Q_3}^2}{2}\left( \Re(\sBot_L)^2+\Re(\sTop_L)^2\right)+\frac{m_{U_3}^2}{2} \Re(\sTop_R)^2+\frac{m_{D_3}^2}{2} \Re(\sBot_R)^2                   \\
	    & +\frac{m_{L_3}^2}{2} \Re(\sTau_L)^2+\frac{m_{E_3}^2}{2} \Re(\sTau_R)^2-m_A^2 \Re(h_d^{0}) \Re(h_u^{0}) \sin (\beta ) \cos (\beta )                     \\
	    & +\frac{1}{2} \Re(h_d^{0})^2 \left(m_{H_d}^2+\mu ^2\right)+\frac{1}{2}\Re(h_u^{0})^2 \left(m_{H_u}^2+\mu ^2\right)\,,                                   \\
	\end{aligned}  \\
	\intertext{the cubic part is given by}
	V_3 & =\begin{aligned}[t]                                                                                                                                    
	    & \frac{y_b}{\sqrt{2}} \Re(\sBot_L) \Re(\sBot_R)\left(A_b  \Re(h_d^{0})-\mu  \Re(h_u^{0})\right)                                                         \\
	    & +\frac{y_\tau}{\sqrt{2}} \Re(\sTau_L) \Re(\sTau_R) \left(A_\tau  \Re(h_d^{0})-\mu  \Re(h_u^{0})\right)                                                 \\
	    & +\frac{y_t}{\sqrt{2}} \Re(\sTop_L)    \Re(\sTop_R) \left( A_t \Re(h_u^{0}) -\mu \Re(h_d^{0}) \right)\,,                                                \\
	\end{aligned}  \\
	\intertext{and the quartic part is given by}
	V_4 & = \begin{aligned}[t]                                                                                                                                   
	    & \frac{g_1^2+g_2^2}{32} \left(\Re(h_u^{0})^2-\Re(h_d^{0})^2\right)^2                                                                                    \\
	    & +\frac{g_1^2-3 g_2^2+12 y_t^2}{48} \Re(h_u^{0})^2 \Re(\sTop_L)^2 +\frac{3 y_t^2-g_1^2}{12} \Re(h_u^{0})^2 \Re(\sTop_R)^2                               \\
	    & +\frac{g_1^2+3 g_2^2}{48} \Re(h_u^{0})^2 \Re(\sBot_L)^2 +\frac{1}{24} g_1^2 \Re(h_u^{0})^2 \Re(\sBot_R)^2                                              \\
	    & +\frac{g_2^2-g_1^2}{16}  \Re(h_u^{0})^2 \Re(\sTau_L)^2 +\frac{1}{8} g_1^2 \Re(h_u^{0})^2 \Re(\sTau_R)^2                                                \\
	    & +\frac{3 g_2^2-g_1^2}{48} \Re(h_d^{0})^2 \Re(\sTop_L)^2 +\frac{g_1^2}{12} \Re(h_d^{0})^2 \Re(\sTop_R)^2                                                \\
	    & -\frac{g_1^2+3 g_2^2-12 y_b^2}{48}  \Re(h_d^{0})^2 \Re(\sBot_L)^2-\frac{g_1^2-6 y_b^2}{24} \Re(h_d^{0})^2 \Re(\sBot_R)^2                               \\
	    & +\frac{g_1^2-g_2^2+4 y_\tau^2}{16} \Re(h_d^{0})^2 \Re(\sTau_L)^2  -\frac{g_1^2-2 y_\tau^2}{8} \Re(h_d^{0})^2 \Re(\sTau_R)^2                            \\
	    & +\frac{g_1^2+9 g_2^2+12 g_3^2}{288} \left(\Re(\sTop_L)^2+\Re(\sBot_L)^2\right)^2-\frac{g_1^2+3 g_2^2}{48} \Re(\sTop_L)^2 \Re(\sTau_L)^2                \\
	    & -\frac{g_1^2+3 g_3^2-9 y_t^2}{36} \Re(\sTop_L)^2 \Re(\sTop_R)^2 +\frac{g_1^2-6 g_3^2+18 y_b^2}{72} \Re(\sTop_L)^2 \Re(\sBot_R)^2                       \\
	    & +\frac{g_1^2}{24} \Re(\sTop_L)^2 \Re(\sTau_R)^2                                                                                                        \\
	    & + \frac{3 g_2^2-g_1^2}{48} \Re(\sBot_L)^2 \Re(\sTau_L)^2 -\frac{g_1^2+3 g_3^2-9 y_t^2}{36} \Re(\sBot_L)^2 \Re(\sTop_R)^2                               \\
	    & + \frac{g_1^2-6 g_3^2+18 y_b^2}{72} \Re(\sBot_L)^2 \Re(\sBot_R)^2   +\frac{g_1^2}{24}  \Re(\sBot_L)^2 \Re(\sTau_R)^2                                   \\
	    & +\frac{g_1^2+g_2^2}{32} \Re(\sTau_L)^4 +\frac{g_1^2}{12}  \Re(\sTau_L)^2 \Re(\sTop_R)^2 -\frac{g_1^2}{24} \Re(\sTau_L)^2  \Re(\sBot_R)^2               \\
	    & -\frac{g_1^2-2 y_\tau^2}{8}\Re(\sTau_L)^2 \Re(\sTau_R)^2                                                                                               \\
	    & +\frac{4 g_1^2+3 g_3^2}{72} \Re(\sTop_R)^4 - \frac{2 g_1^2-3 g_3^2}{36} \Re(\sTop_R)^2 \Re(\sBot_R)^2 -\frac{g_1^2}{6} \Re(\sTop_R)^2   \Re(\sTau_R)^2 \\
	    & +\frac{g_1^2+3 g_3^2}{72}  \Re(\sBot_R)^4 +\frac{g_1^2}{12} \Re(\sBot_R)^2 \Re(\sTau_R)^2 +\frac{g_1^2 }{8} \Re(\sTau_R)^4                             \\
	    & +\frac{y_b y_\tau }{2} \Re(\sBot_L) \Re(\sTau_L) \Re(\sBot_R)  \Re(\sTau_R)\,.                                                                         \\
	\end{aligned}
\end{align}

\section[Constraints from Vacuum Stability in the Other Benchmark Scenarios]{Constraints from Vacuum Stability in the Other Benchmark Scenarios of~\cite{Bahl2018a}}
\label{app:rest}

In this appendix we present our results for the vacuum stability
constraints in the remaining CP-conserving benchmark scenarios defined
in~\cite{Bahl2018a}.

\begin{figure}
	\centering
	\includegraphics[width=\textwidth]{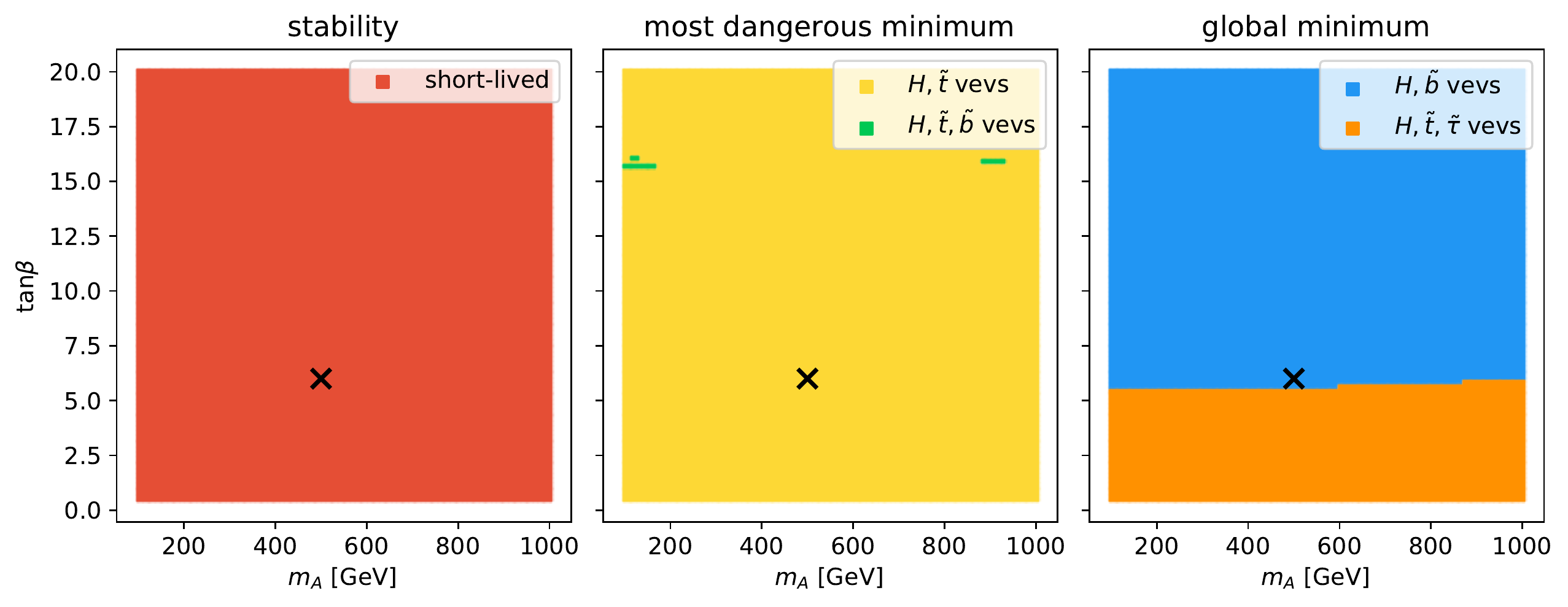}
	\caption{Constraints from vacuum stability in the
		$M_h^{125}(\text{alignment})$ scenario. The left panel
		indicates the lifetime of the EW vacuum, which is short-lived
		all over the displayed plane. In the middle and right panel
		the character of the MDM and the global minimum is
		illustrated, respectively. The black $\boldsymbol{\times}$
		indicates the same point shown in
		\cref{fig:Mh125alitri}.}\label{fig:Mh125ali}
\end{figure}

The vacuum stability analysis for the $M_h^{125}(\text{alignment})$
benchmark plane defined in \cref{eq:parameters_ali} is displayed in
\cref{fig:Mh125ali}. As noted in \cref{sec:ali}, all parameter points
in this plane give rise to short-lived EW vacua. The MDM is characterised
by \sTop{} vevs throughout the parameter plane, while the global minimum
has \sBot{} vevs for $\tan\beta \gtrsim 5.5$ and
\sTop{}-\sTau{} vevs for smaller values of $\tan\beta$.

\begin{figure}
	\centering
	\includegraphics[width=0.35\textwidth]{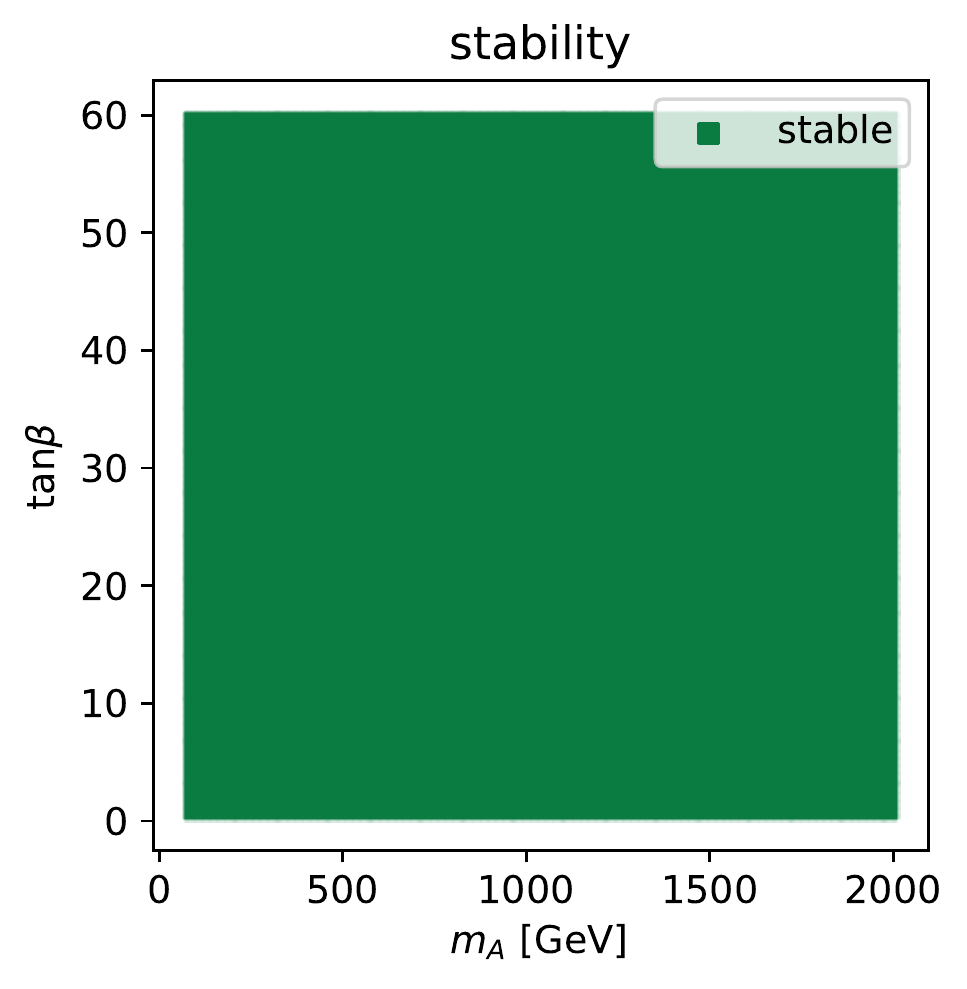}
	\caption{Constraints from vacuum stability in the
		$M_h^{125}(\tilde\chi)$ scenario. The colour code indicates the
		lifetime of the EW vacuum, which is stable throughout the
		displayed plane.}\label{fig:Mh125chi}
\end{figure}

The light neutralino scenario $M_h^{125}(\tilde\chi)$ is defined through
\begin{equation}
	\begin{gathered}
		m_{Q_3}=m_{U_3}=m_{D_3}=1.5\,\TeV\,,\quad m_{L_3}=m_{E_3}=2\,\TeV\,,\quad
		\mu=180\,\GeV\,,\\
		X_t = A_t - \frac{\mu}{\tan\beta}=2.5\,\TeV\,,\quad A_b=A_\tau=A_t\,,\\
		M_1=160\,\GeV\,,\quad M_2=180\,\GeV\,,\quad M_3=2.5\,\TeV\,.
	\end{gathered}
\end{equation}
As shown in \cref{fig:Mh125chi} the entire benchmark plane features an
absolutely stable EW vacuum.  This is unsurprising, since it has a very small
$\mu$ and relatively large soft sfermion masses while the \(A\) parameter is not
much larger than the soft masses. The small gaugino mass parameters do not lead
to instabilities as they only enter the scalar potential through the $\Delta_b$
and $\Delta_\tau$ corrections.

\begin{figure}
	\centering
	\includegraphics[width=\textwidth]{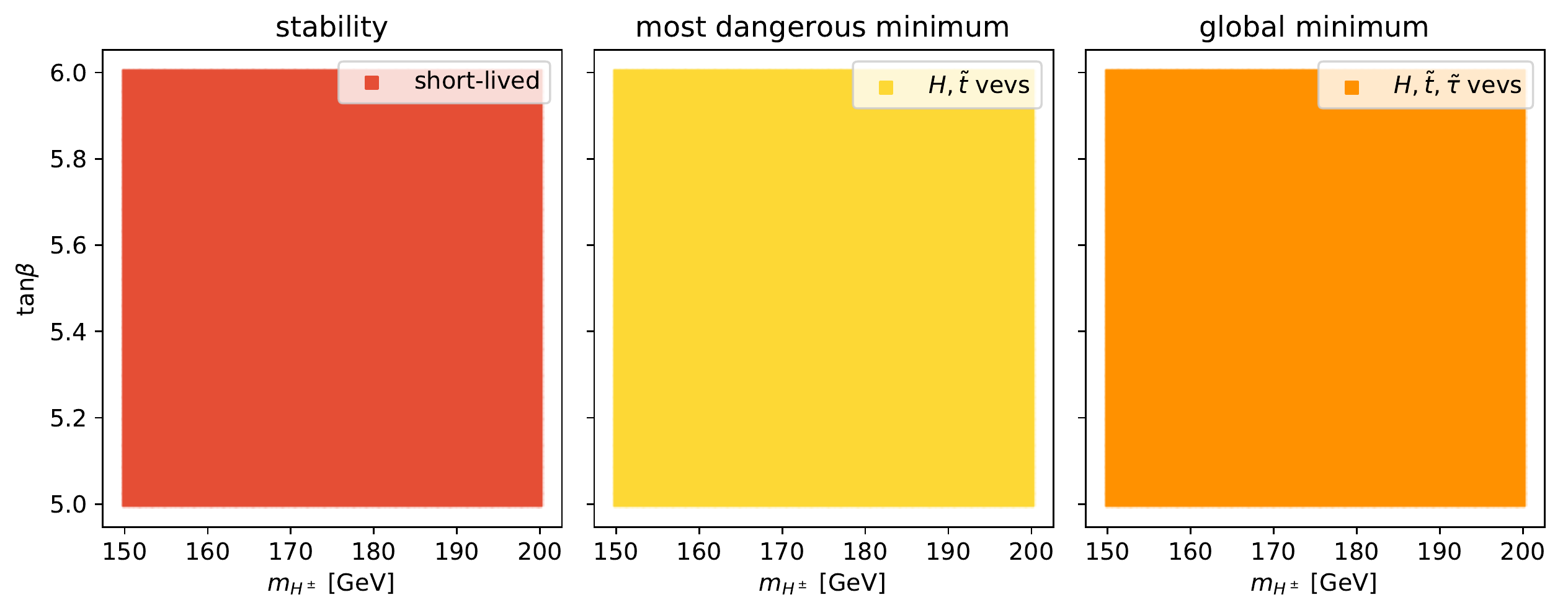}
	\caption{Constraints from vacuum stability in the $M_H^{125}$
		scenario. The left panel indicates the lifetime of the EW
		vacuum, which is short-lived over the displayed plane, while
		the middle and right panels illustrate which fields have
		non-zero vevs at the MDM and the global minimum,
		respectively.}\label{fig:MH125}
\end{figure}

Finally, the heavy Higgs alignment scenario $M_H^{125}$ is defined
through
\begin{equation}
	\begin{gathered}
		m_{Q_3}=m_{U_3}=750\,\GeV-2(m_{H^\pm}-150\,\GeV)\,,\\
		\mu = \left(5800\,\GeV + 20 (m_{H^\pm}-150\,\GeV)\right)m_{Q_3}/(750\,\GeV)\,,\\
		A_t=A_b=A_\tau=0.65 m_{Q_3}\,,\quad m_{D_3}=m_{L_3}=m_{E_3}=2\,\TeV\,,\\
		M_1 = m_{Q_3} - 75\,\GeV\,,\quad M_2 = 1\,\TeV\,,\quad M_3=2.5\,\TeV\,.
	\end{gathered}
\end{equation}
In this scenario the heavy Higgs boson $H$ has a mass of about $125\,\GeV$ with
SM-like couplings. Since the scenario has small soft SUSY-breaking squark mass
parameters and a very large $\mu$ it is unsurprising that the entire benchmark
plane shown in \cref{fig:MH125} has a short-lived EW vacuum. Throughout the
plane, the MDM has non-vanishing \sTop{} vevs while the global minimum has
\sTop{}-\sTau{} vevs. Considering how constrained the region of
phenomenologically viable parameter space for this scenario is (see Fig. 10
of~\cite{Bahl2018a}) we do not expect that a heavy Higgs alignment scenario with
a long-lived vacuum exists in the MSSM\@. However, a detailed assessment of this
possibility would require a dedicated study.

\bibliographystyle{JHEP}
\bibliography{Bibliography}

\end{document}